\documentclass[useAMS,usenatbib]{mn2e}
\usepackage{epsf}
\usepackage{natbib}
\usepackage{epsfig}
\usepackage{subfigure}

\newcommand{\alphaFet}{[$\alpha$/Fe]}
\newcommand{\alphaFe}{[$\alpha$/Fe]\,}
\newcommand{\FeH}{[\rm Fe/H]\,}

\newcommand{\gimic}        {\textsc{gimic}}

\newcommand{\gadget}      {\textsc{gadget-3}}
\newcommand{\subfind}    {\textsc{Subfind}}

\newcommand{\galexev}         {\textsc{Galaxev}}

\title[{Stellar haloes of disc galaxies}]{Cosmological simulations of the formation of the stellar haloes around disc galaxies}

\author[A.~S.~Font et~al.]{A. S. Font$^{1}$\thanks{E-mail:
afont@ast.cam.ac.uk},  I.~G.~McCarthy$^{2,1}$, R.~A.~Crain$^{3}$, T.~Theuns$^{4,5}$, J. Schaye$^{6}$, \newauthor R. P. C. Wiersma$^{6,7}$, C. Dalla Vecchia$^{6,8}$ \ \\
$^{1}$Institute of Astronomy, University of Cambridge, Madingley Road, Cambridge, CB3 0HA\\
$^{2}$Kavli Institute for Cosmology, University of Cambridge, Madingley Road, Cambridge, CB3 OHA\\
$^{3}$Centre for Astrophysics \& Supercomputing, Swinburne University of Technology, Hawthorn, Victoria 3122, Australia\\
$^{4}$Institute of Computational Cosmology, Department of Physics, University of Durham, Science Laboratories, South Road, Durham DH1 3LE\\
$^{5}$Department of Physics, University of Antwerp, Campus Groenenborger, Groenenborgerlaan 171, B-2020 Antwerp, Belgium\\
$^{6}$Leiden Observatory, Leiden University, P. O. Box 9513, 2300 RA Leiden, the Netherlands\\
$^{7}$Max-Planck-Institut f{\"u}r Astrophysik, Karl-Schwarzschild-Stra{\ss}e 1, D-85741 Garching, Germany \\
$^{8}$Max Planck Institute for Extraterrestrial Physics, Giessenbachstra{\ss}e 1, 85748 Garching, Germany
}

\begin{document}

\date{Accepted ... Received ...}

\pagerange{\pageref{firstpage}--\pageref{lastpage}} \pubyear{2008}

\maketitle

\label{firstpage}

\begin{abstract}

We use the Galaxies-Intergalactic Medium Interaction Calculation (\gimic) suite of cosmological hydrodynamical simulations to study
the formation of stellar spheroids of Milky Way-mass disc galaxies.  The simulations contain accurate treatments of metal-dependent radiative cooling, star formation, supernova feedback, and chemodynamics, and the large volumes that have been simulated yield an unprecedentedly large sample of $\approx400$ simulated $\sim L_*$ disc galaxies.  The simulated galaxies are surrounded by low-mass, low-surface brightness stellar haloes that extend out to $\sim 100$ kpc and beyond.  The diffuse stellar distributions bear a remarkable resemblance to those observed around the Milky Way, M31 and other nearby galaxies, in terms of mass density, surface brightness, and metallicity profiles.  We show that in situ star formation typically dominates the stellar spheroids by mass at radii of $r \la 30$~kpc, whereas accretion of stars dominates at larger radii and this change in origin induces  a change in slope of the surface brightness and metallicity profiles, which is also present in the observational data.  The system-to-system scatter in the in situ mass fractions of the spheroid, however, is large and spans over a factor of $4$.  Consequently, there is a large degree of scatter in the shape and normalisation of the spheroid density profile within $r \la 30$~kpc (e.g., when fit by a spherical powerlaw profile the indices range from $-2.6$ to $-3.4$).  We show that the in situ mass fraction of the spheroid is linked to the formation epoch of the system.  Dynamically older systems have, on average, larger contributions from in situ star formation, although there is significant system-to-system scatter in this relationship.
Thus, in situ star formation likely represents the solution to the longstanding failure of pure accretion-based models to reproduce the observed properties of the inner spheroid.  

\end{abstract}


\section{Introduction}
\label{sec:intro}

The stellar haloes of normal disc galaxies are expected to be excellent repositories of fossil evidence from the epoch of galaxy formation, even though these components contain only a small fraction of the total light.  This is because the stellar haloes are widely believed to have been formed through the disruption of infalling satellite dwarf galaxies over the lifetime of the galaxy.  
Recent advances in observational techniques have allowed the detection of faint stellar haloes around external galaxies, such as M31 \citep{ferguson02,guhathakurta05,irwin05,ibata07} and other nearby disc galaxies \citep{mouhcine05a,mouhcine05b,dejong07,ibata09,mouhcine10}. Most stellar haloes have been shown to contain evidence of merger events in the form of tidal stellar streams. The number and physical properties of these streams appear to be in general agreement with the predictions of the $\Lambda$CDM model \citep{bell08,gilbert09a,gilbert09b,mcconnachie09,starkenburg09,martinez-delgado10}. 

In spite of this, questions have been raised about the compatibility with the standard paradigm of the seemingly large differences in the properties of the stellar haloes and satellite systems of M31 and the Milky Way.  These two systems have similar masses but M31 appears to have experienced a much more active merger history in the recent past, as suggested by its more disturbed stellar disc, larger bulge, younger and more metal-rich stellar populations, larger and more numerous tidal streams and surviving satellite galaxies \citep{guhathakurta05,irwin05,brown06a,brown06b,gilbert06,kalirai06,mcconnachie06,gilbert07,ibata07,koch08,kalirai09}. 

On the theoretical side, the detection of extended stellar components around external galaxies  has reinvigorated the efforts of modelling stellar haloes in the context of $\Lambda$CDM cosmogony. Several theoretical studies suggest that the main properties of stellar haloes can be explained entirely by the accretion and disruption of satellites (\citealt{bullock05,font06a,font06b,font06c,delucia08,font08,johnston08,gilbert09b,cooper10}). These studies have been carried out using a combination of high-resolution dark matter-only simulations and simple semi-analytic prescriptions for assigning light (i.e., stellar mass) to infalling satellites.  Recent studies using such `hybrid' methods (e.g., \citealt{cooper10}) have taken advantage of the ultra-high resolution achieved by the latest generation of dark matter only Milky Way simulations (\citealt{springel08}; see also \citealt{diemand07}), which allowed them to explore the properties of even the faintest substructures (satellites and streams) surviving to the present day.

However, even though a large fraction of the stellar halo\footnote{By volume, but not by mass as we argue later in this paper.} of normal disc galaxies may be explained by the accretion scenario, there remain some stellar populations in the Milky Way and other nearby galaxies that have not so far been accounted for in the context of accretion-only models.  This has been known for a long time \citep{eggen62}, but the evidence has become more convincing in the past few years. For example, the metal-rich halo stars in the Solar neighbourhood have been shown to have a bi-modal \alphaFe pattern \citep{nissen10}. In particular, the high \FeH - high \alphaFe population seems to be the product of very rapid chemical evolution that cannot be sustained in the potential wells of even the most massive accreted satellites \citep{font06a}. Instead, it has been suggested that these stars may have formed ``in situ'' in the halo (e.g., \citealt{zolotov10}) or they may have been stirred up from the disc by repeated interactions with satellite galaxies \citep{purcell10}. Similarly, models that track only the accreted component of the stellar halo (e.g., \citealt{font08}) have difficulties matching the intermediate age ($\sim 8$~Gyr), metal-rich populations (\FeH $> -1$) detected in the halo of M31 (see \citealt{brown07}). 

There may also be an indication of a spatial differentiation in the stellar halo of the Milky Way: analysis of main-sequence turnoff stars in the Sloan Digital Sky Survey (SDSS) suggests that the inner regions are more metal-rich (e.g., \citealt{carollo07,carollo10,dejong10}, but see \citealt{sesar11}) and have a prograde motion with respect to the disc, while the outer halo is metal-poor and appears to have a small net retrograde rotation or no rotation, depending on the adopted the local standard of rest \citep{deason11a} and/or the adopted distance scale \citep{schoenrich10,beers11}.  In addition, there is mounting evidence for a break in the slope of the density profile of the halo beyond $r \sim 25$ kpc \citep{sesar11,deason11b}.  Furthermore, recent studies suggest that the inner regions of the halo are flattened (e.g., \citealt{juric08,sesar11,deason11b}), which is similar to what is seen for the inner regions of M31 (e.g., \citealt{pritchet94}).  These results suggest that the stellar haloes of galaxies may have a dual nature, perhaps along the lines of the scenario proposed originally by \citet{searle78}: while the outer halo formed primarily by accretion/mergers, some of the inner halo has formed through dissipational collapse.  However, in the current paradigm, the dissipative component would not have formed through monolithic collapse \citep{eggen62}, but rather as a result of gas-rich mergers at early times.  Gas-dynamical simulations of disc galaxies formed in a hierarchical scenario produce stellar systems with a significant in situ dissipative component \citep{abadi06,brook04,zolotov09,zolotov10,oser10}.  It remains to be determined what the relative contributions of in situ and accreted stars are as a function of radius and system mass and, importantly, what the system-to-system scatter is in these contributions.  The scatter in the in situ and accreted components may be at the heart of the reported large differences in the properties of the extended stellar distributions of the Milky Way and M31.

Another important aspect that we address here, which may be related to the in situ vs.\ accretion question, is the origin of large-scale metallicity gradients in stellar haloes.  The emergence of metallicity gradients is a well-known consequence of dissipative collapse/mergers.  The evidence in favour of metallicity gradients in local disc galaxies is mounting.  For example, in the Milky Way, an underlying metallicity gradient in the halo is suggested by the net difference in the mean metallicity between the inner and outer halo (\citealt{eggen62,searle78,carollo07,carollo10,dejong10}, but see \citealt{sesar11}). The halo of M31 also shows evidence for a gradient, where the inferred metallicity may drop by as much as $1$~dex from the inner regions of the galaxy out to $\approx100$~kpc (see Fig.~\ref{fig:iron_profs} below). NGC 891 represents another local Milky Way-analog for which there is evidence of a stellar metallicity gradient (see \citealt{ibata09}; although note that metallicity measurements have only been made out to 15 kpc for this system). 

Models that form haloes by non-dissipative accretion and mergers have so far not been able to produce significant metallicity gradients \citep{font06a,delucia08,cooper10}.  This suggests that either these models do not capture all the relevant physical processes (e.g., in situ star formation or ejection of disc stars by interaction with satellites) or that they do not model the entire range of possible merger histories for Milky Way-mass galaxies in the cosmological context. In this study we use cosmological hydrodynamical simulations with a sophisticated treatment of chemodynamics to show that metallicity gradients are expected to be ubiquitous in the haloes of Milky Way-mass disc galaxies.  The prominence of these gradients is shown to be tied to the fraction of the stellar mass that is formed in situ, which is itself determined by the merger histories of the galaxies. 

The paper is organised as follows. In Section \ref{sec:sims} we describe the simulations and the selection of the sample Milky Way-mass disc galaxies. In Section \ref{sec:halo} we compute the average properties of stellar haloes of these galaxies, such as their density profiles, metallicity gradients (both \FeH and \alphaFet) and metallicity distribution functions (MDFs). In Section \ref{sec:origin} we quantify the contribution to the stellar spheroid from in situ star formation and accretion/disruption of satellites.  Finally, we summarize and discuss our main findings in Section \ref{sec:concl}.

\section{Simulations}
\label{sec:sims}

We use the Galaxies-Intergalactic Medium Interaction Calculation (\gimic) suite of cosmological hydrodynamical simulations, which were carried out by the Virgo Consortium and are described in detail by \citet{crain09}, hereafter C09 (see also \citealt{crain10}, hereafter C10).  We will therefore present only a brief summary of the simulations here, focusing on the details most relevant for the present study.  This suite of simulations is ideal for the present study for several reasons: (i) it simulates large volumes of the universe that contain numerous Milky Way-mass disc galaxies, allowing us to robustly quantify both the mean trends and the scatter; (ii) it incorporates accurate physical prescriptions for metal-dependent radiative cooling, star formation, mass and energy feedback from Type Ia and Type II supernovae, as well as enrichment due to stellar evolution (AGB stars), which results in the formation of realistic disc galaxies; and (iii) the relatively high numerical resolution is adequate for resolving the stellar haloes of individual Milky Way-mass galaxies and their most massive satellite galaxies (i.e, the main contributors to the accreted component of the stellar halo).

The suite consists of a set of hydrodynamical re-simulations of five nearly spherical regions ($\sim 20 h^{-1}$ Mpc in radius) extracted from the Millennium Simulation \citep{springel_etal_05a}.  The regions were selected to have overdensities at $z=1.5$ that represent $(+2, +1, 0, -1, -2) \sigma$, where $\sigma$ is the root-mean-square deviation from the mean on this spatial scale.  The 5 spheres are therefore environmentally diverse in terms of the large-scale structure that is present within them.  For the purposes of the present study, however, we will only select systems with total `main halo' (i.e., the dominant subhalo in a friends-of-friends, hereafter FoF, group) masses similar to that of the Milky Way, irrespective of which \gimic\ volume it is in (i.e., irrespective of the large-scale environment).  C09 found that the properties of systems of fixed main halo mass do not depend significantly on the large-scale environment (see, e.g., Fig.\ 8 of that study).  We plan to study the properties of such galaxies as a function of their {\it local} environment in a future study.

The cosmological parameters adopted for \gimic\ are the same as those assumed in the Millennium Simulation and correspond to a $\Lambda$CDM model with $\Omega_{m} = 0.25$, $\Omega_{\Lambda} = 0.75$, $\Omega_{b} = 0.045$, $\sigma_{8} = 0.9$ (where $\sigma_{8}$ is the rms amplitude of linearly evolved mass fluctuations on a scale of $8 h^{-1}$ Mpc at $z=0$), $H_{0} = 100 h$ km  s$^{-1}$ Mpc$^{-1}$, $h = 0.73$, $n_s=1$ (where $n_s$ is the spectral index of the primordial power spectrum).  The value of $\sigma_8$ is approximately 2-sigma higher than has been inferred from the most recent CMB data \citep{komatsu09}, which will affect the abundance of Milky Way-mass systems somewhat, but should not significantly affect their individual properties.

The simulations were evolved to $z=0$ using the TreePM-SPH code \gadget\ 
(last described in \citealt{springel05}). The \gadget\ code has been 
substantially modified to incorporate new baryonic physics. Radiative cooling 
rates for the gas are computed on an element-by-element basis by interpolating 
within pre-computed tables generated with CLOUDY \citep{ferland98} that contain 
cooling rates as a function of density, temperature, and redshift and that 
account for the presence of the cosmic microwave background and 
photoionisation from a \citet{haardt01} ionising UV/X-Ray background 
(see \citealt{wiersma09a}).  This background is switched on at $z=9$ in the 
simulation, where it `reionises' the entire simulation volume.  Star formation 
is tracked in the simulations following the prescription of \citet{schaye08}.   
Gas with densities exceeding the critical density for the onset of the thermo-gravitational 
instability is expected to be multiphase and to form stars \citep{schaye04}. Because the
simulations lack both the physics and the resolution to model the cold interstellar gas 
phase, an effective equation of state (EOS) is imposed with pressure $P \propto \rho^{4/3}$  
for densities\footnote{Gas particles are only placed on the EOS if their temperature is below $10^5$ K when they cross the density threshold and if their density exceeds 57.7 times the cosmic mean.  These criteria prevent star formation in intracluster gas and in intergalactic gas at very high redshift, respectively \citep{schaye08}.} $n_H > n_*$ where $n_* = 0.1$ cm$^{-3}$.  As described in \citet{schaye08}, gas on the effective EOS is allowed to form stars at a pressure-dependent
rate that reproduces the observed Kennicutt-Schmidt law \citep{kennicutt98} by construction.  
The timed release of individual elements by both massive (Type II SNe and 
stellar winds) and intermediate mass stars (Type Ia SNe and asymptotic giant branch stars) 
is included following the prescription of \citet{wiersma09b}.  A set of 11
individual elements are followed in these simulations (H, He, C, Ca, N, O, Ne, Mg, S, Si, 
Fe), which represent all the important species for computing radiative cooling rates.

Feedback from supernovae is incorporated using the kinetic wind model 
of \citet{dallavecchia08} with the initial wind velocity, $v_w$, set to $
600$ km/s and the mass-loading parameter (i.e., the ratio of the mass of gas 
given a velocity kick to that turned into newly formed star particles), 
$\eta$, set to $4$.  This corresponds to using 
approximately 80\% of the total energy available from supernovae for a \citet{chabrier03} IMF, which is assumed in the simulation.  This choice of parameters results in a good match to the peak of the star formation rate history of the universe (C09; see also \citealt{schaye10}) and reproduces a number of X-ray/optical scaling relations for normal disc galaxies (C10).

The \gimic\ simulations have been run at three levels of numerical resolution, `low', `intermediate', and `high', to test for numerical convergence.  The low-resolution simulations have the same mass resolution as the Millennium Simulation (which, however, contains only dark matter) while the intermediate- and high-resolution simulations have $8$ and $64$ times better mass resolution, respectively.  In the case of the high-resolution simulations, only the $-2\sigma$ volume was run to $z=0$, owing to the computational expense of running such large volumes at this resolution.  For the purposes of the present study, we therefore follow the approach of C10 and adopt the intermediate-resolution simulations in the main analysis and reserve the high-resolution $-2\sigma$ simulation to test the numerical convergence of our results.

\begin{figure*}
\includegraphics[width=18cm]{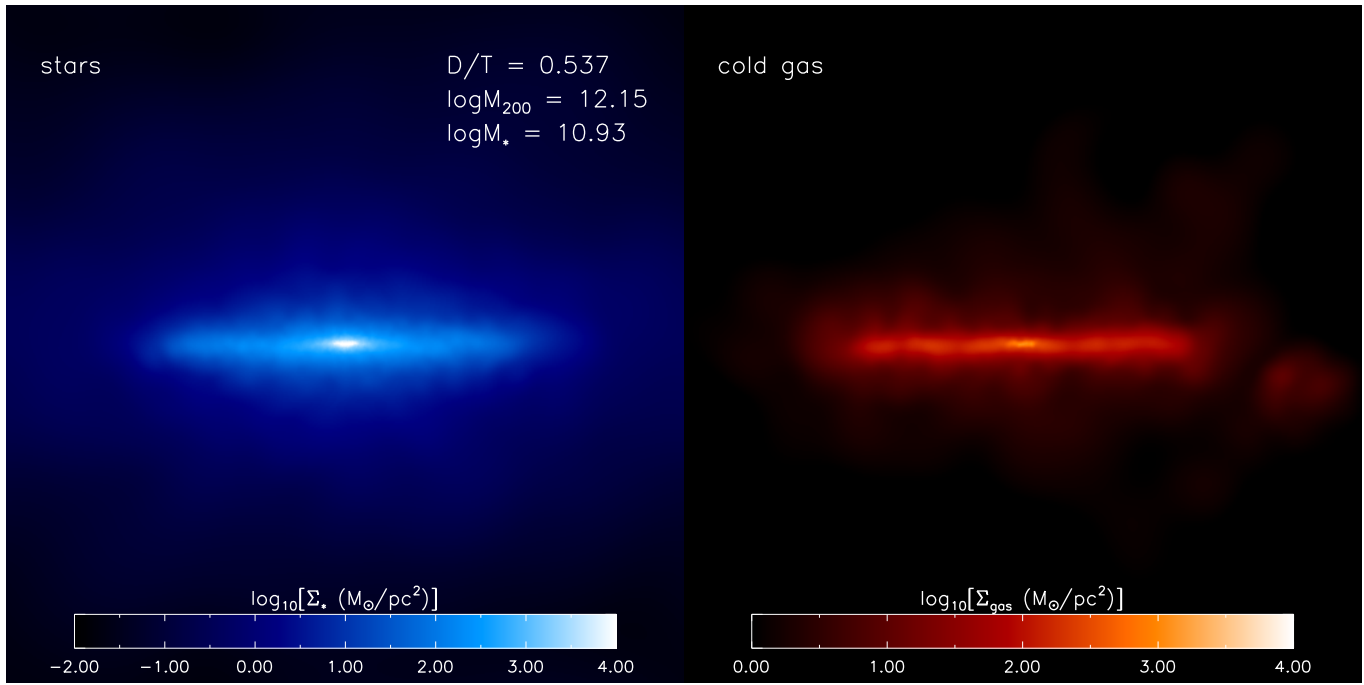}
\caption{\label{fig:galaxy} Surface mass density maps of all the stars ({\it left}) and gas ({\it right}) in one of the sample Milky Way-mass disc galaxies at $z=0$ in the high-resolution simulation. The panel is 100~kpc on a side.}
\end{figure*}

The intermediate-resolution runs therefore have a dark matter particle mass $m_{\rm DM} \simeq 5.30 \times 10^{7} h^{-1}$ M$_{\odot}$ and an initial gas particle mass of $m_{\rm gas} \simeq 1.16 \times 10^{7} h^{-1}$ M$_{\odot}$, implying that it is possible to resolve satellites with total stellar masses similar to those of the classical dwarf galaxies around the Milky Way (with $M_{*} \sim 10^{9-10} M_{\odot}$) with several hundred up to $\sim1000$ particles.  This is important for the present study, as it is currently believed that the disruption of such massive satellites is the primary contributor by mass to the {\it accreted} stellar halo \citep{font06a,cooper10}.  Without a sufficiently large number of particles, the internal structure of the satellites will not be adequately resolved which can lead to overefficient tidal stripping of the satellites.  In addition, relatively high resolution is required to model star formation in the satellites correctly.  For example, \citet{brooks07} and \citet{christensen10} find they require at least several thousand to $\sim10^4$ particles before convergent star formation histories are obtained for isolated dwarf galaxies simulated with the SPH code {\textsc{gasoline}}.  This is similar to the number of particles with which we resolve the most massive satellites galaxies in the high-resolution \gimic\ run.  In the Appendix, we present a convergence study for our simulations, comparing the results of the intermediate- and high-resolution \gimic\ simulations.  We show that the present-day luminosity function of satellites, as well as the mass and metal distribution of the stellar halo, in our default intermediate-resolution runs are robust to an increase in the mass resolution of a factor of 8.  In Table~\ref{tab:res} we present a summary of the mass and force resolution of the simulations used in this study.

\begin{table}
\caption{Resolution of the \gimic\ runs.  $N_{\rm DM}$ and $N_*$ are the median number of dark matter and star particles within $r_{200}$ of the simulated galaxies.  $m_{\rm DM}$ and $m_{\rm gas}$ are the dark matter and (initial) gas particle masses.  $\epsilon_{\rm soft}$ is the Plummer equivalent force resolution in physical space at $z \le 3$.}
\centering
\begin{tabular}{cccccc} \hline
Run & $N_{\rm DM}$ & $N_*$ & $m_{\rm DM}$ & $m_{\rm gas}$ & $\epsilon_{\rm soft}$ \\
  & ($10^4$) & ($10^4$) & ($M_\odot/h$) & ($M_\odot/h$) & (kpc$/h$)\\
\hline
Int-res & $1.6$ & $0.59$ & $5.30\times10^7$ & $1.17\times10^7$ & 1.0\\
Hi-res  & $9.1$ & $3.7$ & $6.63\times10^6$ & $1.46\times10^6$ & 0.5\\
\hline
\end{tabular}
\label{tab:res}
\end{table} 


Note that for all these simulations the remainder of the ($500 h^{-1}$ Mpc)$^{3}$ Millennium volume is also simulated, but with dark matter only and at much lower resolution.  This ensures that the influence of the surrounding large-scale structure is accurately accounted for.

\subsection{Selection of Milky Way-mass systems}
\label{sec:sample}

Below we describe the selection criteria for our sample of Milky Way-mass systems.  

First we define what we mean by a `Milky Way-mass' system.  This is a system with a present-day total (gas+stars+dark matter) mass within $r_{200}$ (i.e., the radius which encloses a mean density of 200 times the critical density of the universe at $z=0$) of $7\times10^{11} M_\odot < M_{200} < 3\times10^{12} M_\odot$.  For reference, this range roughly spans the mass estimates in the literature for both the Milky Way and M31 \citep{kochanek96,evans00,battaglia05,karanchentsev06,guo10}, although we are not attempting to reproduce the properties of either of these systems in detail.  Systems and their substructures are identified in the simulations using the \subfind\ algorithm of \citet{dolag09}, that extends the standard implementation of \citet{springel01} by including baryonic particles in the identification of self-bound substructures.  For each FoF system that is identified, a spherical overdensity (SO) mass with $\Delta = 200$ is computed (i.e., $M_{200}$).  The gas+stars+dark matter associated with the most massive subhalo of a FoF system are considered to belong to the `galaxy', while the gas+stars+dark matter associated with the other substructures are classified as belonging to satellite galaxies.  As the current study is focused on examining the extended stellar distributions of normal disc galaxies, we exclude all bound substructures (i.e., satellites) from the analyses presented in this paper, with the exception of Fig.~\ref{fig:sat_LF} (described below).

Galaxies identified in this way are then classified morphologically as disc- or spheroid-dominated, based on their dynamics.  In particular, we use the ratio of disc stellar mass to total stellar mass (D/T) within 20 kpc computed by C10.  We select disc galaxies with $D/T > 0.3$, although very similar results are obtained for other cuts in D/T (e.g., we have also tried 0.2 and 0.4 and none of the main results of our study are changed).  C10 decomposed the spheroid component from the disc by first computing the angular momentum vector of all the stars within 20 kpc, calculating the mass of stars that are counter-rotating, and doubling it.  This procedure therefore assumes the spheroid has no net angular momentum.  The remaining mass of stars is assigned to the disc and D/T is computed as the disc mass divided by the sum of the disc and spheroid masses.  C10 plotted the D/T distribution in their Fig.\ 1 for a sample of galaxies in the intermediate resolution \gimic\ runs with a very similar mass range to the one adopted in the present study.  They found that approximately half of all the simulated Milky Way-mass galaxies are disc-dominated (D/T $> 0.5$), while approxiately $20\%$ lack a significant spheroidal component D/T $> 0.75$ (i.e., are effectively ``bulgeless'', Sales et al. in prep).

The method of C10 for decomposing the disc and spheroid does not assign individual star particles to either component, it simply computes the {\it total} mass in two components.  We therefore follow the approach of \citet{abadi03} in assigning particles to the disc and spheroid.  In particular, we align the total angular momentum of the stellar disc with the z-axis, calculate the angular momentum of each star particle in the x-y plane, $J_z$, and compare it with the angular momentum, $J_{\rm circ}$, of a star particle on a co-rotating circular orbit with the same energy.  Disc particles are selected using a cut of $J_z/J_{\rm circ} > 0.8$ as in \citet{zolotov09}.  Visually, this cut is quite successful in isolating the disc, but in a small number of cases we found that particles well above/below the disc mid-plane were also assigned to the disc.  For this reason, we also apply a spatial cut such that disc particles cannot exist more than two softening lengths above or below the disc mid-plane. Note that we have also tried assigning disc particles based on the fraction of stellar kinetic energy in ordered rotation, as defined in \citet{sales10}, but found no significant differences from the method outlined above.

One can also compute the D/T ratio using the method of \citet{abadi03} with the angular momentum cut of \citet{zolotov09}.  Comparison of the D/T values computed this way to that of the method of C10 shows a strong correlation, in the sense that both methods agree on which galaxies are most `disky'.  In general, though, the D/T ratio computed using the method of C10 is $\approx 0.2$ larger on average than that computed using the method of \citet{abadi03} with the angular momentum cut of \citet{zolotov09}.  The implication of this is that the spheroid does have net angular momentum in reality (or else that the stellar disc has a much broader range of angular momenta than that advocated by \citealt{zolotov09}).  As we have noted above, however, our main conclusions are insensitive to the exact cut in D/T adopted for our galaxy sample selection.

\begin{table*}
\caption{Global properties of the simulated disc galaxies.  Presented is the median of the property in 3 different total mass ($M_{200}$) bins, with the error bars representing the $5^{\rm th}$ and $95^{\rm th}$ percentiles.  $M_V$ is total V-band magnitude, $M_*(< r_{200})$ is the total stellar mass within $r_{200}$, $v_{\rm rot}(R_\odot)$ is the rotation velocity of the stellar disc at the Solar radius, [Fe/H]$_{r < 30 {\rm kpc}}$ is the mean stellar metallicity within 30 kpc, [Fe/H]$_{r > 30 {\rm kpc}}$ is the mean stellar metallicity beyond 30 kpc, and $n_{\rm gal,bin}$ is the number of simulated galaxies in the total mass bin.}
\centering
\begin{tabular}{ccccccc} \hline
log $M_{200}$ bin & $M_V$ & $M_{*}(< r_{200})$ & $v_{\rm rot}(R_\odot)$ & [Fe/H]$_{r < 30 {\rm kpc}}$ & [Fe/H]$_{r > 30 {\rm kpc}}$ & $n_{\rm gal,bin}$\\
$(M_\odot)$ & (mag.) & $(10^{10} M_\odot)$ & (km/s) & & \\
\hline
$11.85-12.05$ & $-21.37^{-0.98}_{+1.46}$ & $3.46^{+6.32}_{-2.15}$ & $184^{+76}_{-52}$ & $-0.49^{+0.27}_{-0.28}$ & $-1.13^{+0.27}_{-0.19}$ & 127\\
& \\
$12.05-12.25$ & $-22.17^{-0.47}_{+1.53}$ & $8.18^{+5.66}_{-6.32}$ & $243^{+58}_{-93}$ & $-0.35^{+0.18}_{-0.37}$ & $-1.12^{+0.16}_{-0.17}$ & 154 \\
& \\
$12.25-12.50$ & $-22.45^{-0.50}_{+0.69}$ & $12.82^{+8.83}_{-8.85}$ & $280^{+85}_{-89}$ & $-0.30^{+0.13}_{-0.25}$ & $-1.15^{+0.23}_{-0.18}$ & 128\\
\hline
\end{tabular}
\label{tab:global_props}
\end{table*} 

Unlike \citet{zolotov09}, we have elected not to further decompose the spheroid into bulge and halo components.  While separating the disc from the spheroid is relatively straightforward to do for both the observations and the simulations, separating the bulge from the stellar halo is more challenging, as there is no sharp distinction in the dynamics or the spatial distribution of the two components.  A variety of techniques have been used in theoretical and observational studies to distinguish these components, which can complicate the comparison of any single component between different studies (at least in the inner regions where the bulge and halo overlap spatially).  Our approach is to avoid this complication by simply reporting on the total bulge+halo component, which can readily be deduced from observations. 

Lastly, we point out that we have not imposed any explicit constraints on the merger histories of the simulated galaxies.  This is in contrast with many of the previous cosmological modelling studies of Milky Way-type galaxies, which usually select their simulation candidates from systems that have not undergone any major mergers in the recent past (e.g., \citealt{diemand05,okamoto05,cooper10}).  This condition is meant to ensure that these systems develop stellar discs that survive until the present day and that resemble the Milky Way's disc \citep{toth92}.  Note that these previous studies were based on hybrid methods applied to dark matter-only simulations and therefore did not have the benefit of being able to check which histories actually give rise to discs at the present day.  Importantly, more than $70\%$ of Milky Way-mass haloes have experienced major mergers (defined as $M_{\rm sat}/M_{\rm disc} \ge 1/3$, where $M_{\rm sat}$ is the total mass of the satellite and $M_{\rm disc}$ is the stellar mass of the disc) in the last 10 Gyr \citep{stewart08,boylan-kolchin10}, implying that the majority of merger histories may be excluded from studies that select only quiescent histories.  The constraints applied to previous hybrid studies appear to be overly conservative, given that the majority of the Milky Way-mass systems in \gimic\ are disc-dominated systems (see C10).  The resiliency of galactic discs to mergers likely stems from the fact that the galaxies contain a significant dissipational gaseous component (that was not taken into account in, e.g., \citealt{toth92}), which is able to maintain (or re-establish) the disc (see \citealt{hopkins09}).  In any case, for the present study, which is based on hydrodynamic simulations of large volumes, there is no need to place explicit constraints on the merger histories since we know which galaxies have prominent discs at $z=0$.

Applying the system mass and morphology criteria described above yields a total sample of 412 systems in the 5 intermediate-resolution \gimic\ volumes.  This greatly exceeds the sample size of all previous studies of this type (i.e., those based on hydrodynamic re-simulations or hybrid methods), albeit at lower resolution than some previous studies (such as \citealt{zolotov09}).  The large number of systems in our sample allows us to robustly quantify the system-to-system scatter present in the properties of stellar haloes.  Our work is therefore complementary to studies such as \citet{abadi06} and \citet{zolotov09}, which are based on high-resolution simulations of a small number of systems.  (But note we demonstrate in the Appendix that the properties of our simulated stellar haloes are robust to increased numerical resolution.)

As an example, Fig.~\ref{fig:galaxy} shows the stellar and cold gas mass distributions of a typical simulated disc galaxy.  An extended, flattened stellar distribution is clearly visible out to large radii.

\section{Properties of Stellar Haloes}
\label{sec:halo}

Before analysing the detailed properties of the stellar haloes of the simulated disc galaxies, we briefly discuss the overall global properties of the sample galaxies.  The motivation for this is simple: if the simulated galaxies were to look wholly unlike observed normal disc galaxies, then that would limit what can be concluded about the formation of their stellar haloes.  As we show in Section 3.1, the simulated galaxies have global properties (i.e., total luminosities and stellar masses, rotation velocities, and satellite luminosity functions) which are comparable to those estimated for the Milky Way and M31.  The reader who is interested only in the properties of the stellar haloes of the simulated galaxies may wish to skip ahead to Section 3.2.

\subsection{Global properties}
\label{sec:global}

In Table 1 we present the typical properties of the simulated disc galaxies, split into several halo mass ($M_{200}$) bins.  In particular, we list total V-band magnitude ($M_V$, in AB mags.), total stellar mass\footnote{When referring to stellar masses throughout the paper we are always referring to the {\it current} stellar mass, which is less than the initial stellar mass due to stellar mass loss.  Typically, about 45\% of the initial mass is lost due to stellar evolution, which is dictated by the Chabrier IMF adopted in the simulations and the generally old stellar populations we are studying here.} within $r_{200}$ [$M_{*}(< r_{200})$], disc rotation speed at the solar radius [$v_{rot}(R_\odot)$, where $R_\odot$ is assumed to be $8.5$~kpc] and mean stellar metallicity within and beyond a galacto-centric distance of 30 kpc ([Fe/H]$_{r < 30 {\rm kpc}}$).  (The choice of 30 kpc is motivated below, where we show how the physical nature of the bulge+halo component changes with radius.)  The values listed in Table 1 are medians for galaxies in the mass bin, while the error bars represent the $5^{\rm th}$ and $95^{\rm th}$ percentiles.  We also list the number of galaxies in each mass bin ($n_{\rm gal,bin}$). 

The V-band luminosity of star particles in the simulations is calculated as follows.  Each star particle is treated as a simple stellar population (SSP). The ages and metallicities of star particles are used to compute the spectral energy distribution (SED) by interpolating over the \galexev\ models of \citet{bruzual03}.  The V-band luminosity is obtained by integrating the product of the SED with the V-band filter transmission function.  We ignore the effects of dust attenuation in this calculation.

As already noted, the \gimic\ simulations successfully reproduce the peak of the observed star formation rate density history of the Universe, as well as the X-ray-optical scaling properties of normal disc galaxies (C10).  Another notable success of the simulations is that they produce large numbers of disc galaxies (see Fig.\ 1 of C10).  Realistic spiral discs have been notoriously difficult to achieve in the past with cosmological hydrodynamic simulations.  We will present a detailed analysis of the galaxy discs in a future study (McCarthy et al., in prep.).  

The \gimic\ simulations, however, do not reproduce the bright end of the global galaxy luminosity function, $M_{V} < -22$ (see C09). This is perhaps not surprising, as the simulations lack feedback from supermassive black holes, which is thought to be crucial for regulating star formation in the most massive systems (e.g., \citealt{springel_etal_05b,bower06,croton06,booth09}).  The match to the observed abundance of $\sim L_*$ galaxies is, however, reasonable.  One should also recall that the luminosity function depends not only on the luminosity as a function of halo mass (which we quote below), but also on the number density of dark matter haloes, which depends on the adopted cosmological parameters.  This implies that one can obtain correct properties of individual galaxies but still not reproduce the observed luminosity function, if the wrong cosmological parameters are adopted.  Therefore, a better approach to assessing whether the simulated galaxies are realistic, is to compare the mass-to-light ratio (or total luminosity) as a function of halo mass with the observations, which we do immediately below.

\begin{figure}
\includegraphics[width=\columnwidth]{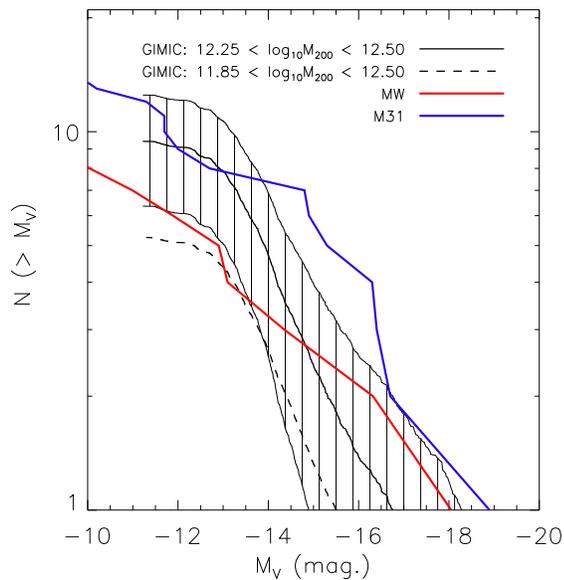} 
\caption{\label{fig:sat_LF} The mean satellite luminosity function of simulated galaxies in the intermediate-resolution \gimic\ simulation.  The black dashed curve represents the mean luminosity function for all the simulated galaxies.  The thick black solid curve represents the mean luminosity function for simulated galaxies in the mass range $1.8\times10^{12} \la M_{200}/M_\odot \la 3.2\times10^{12}$ (128 disc galaxies meet this criteria). The hatched region represents the Poisson error distribution for this luminosity function.  The solid, red and blue histograms show the satellite luminosity functions for the Milky Way and M31, respectively.  The Milky Way data is taken from \citet{mateo98} and the M31 data is taken from \citet{mcconnachie09}.  Overall, massive galaxies in our sample produce luminosity functions that are quite similar to those observed for the Milky Way and M31, suggesting that the accreted component of the halo in the simulations should have a realistic mass fraction.}
\end{figure}

For comparison, M31 has an absolute visual magnitude of $M_V \approx -21.0$, $M_* \approx 5.6 \times 10^{10} M_\odot$, and $v_{\rm rot} \approx 250$ km/s.  The absolute magnitude was calculated using the {\it GALEX} apparent V magnitude of \citet{gildepaz07} and assuming a distance to M31 of 785 kpc.  The total stellar mass was calculated using B-V and V of \citet{gildepaz07} to infer the stellar $M/L$ using the relation of \citet{bell01} and scaling the result by 0.7 to convert from the Salpeter IMF assumed by \citet{bell01} to a Chabrier IMF, which is adopted in the \gimic\ simulations.  The rotation velocity, which has been corrected for inclination, was obtained from the {\small HyperLeda} data base\footnote{http://leda.univ-lyon1.fr.}.  The Milky Way has a total stellar mass of $M_* \approx 5.5 \times 10^{10} M_\odot$ \citep{flynn06}\footnote{\citet{flynn06} directly measure the stellar $M/L$ locally and apply it to the Galaxy to estimate the total stellar mass.  The measured value of $M/L$ is consistent with the predictions of a Chabrier IMF.} and $v_{\rm rot} = 250$ km/s \citep{reid09}.  These properties of M31 and the Milky Way are quite typical of the simulated galaxies, particularly those with halo masses of $M_{200} \la 2\times10^{12}M_\odot$.  We will comment in detail on the metallicity of the stellar halo of both observed and simulated systems below.

How do the D/T ratios for M31 and the Milky Way compare with that of our simulated galaxies?  The D/T ratio for M31 and the Milky Way have been estimated by \citet{geehan06} and \citet{dehnen98}, respectively, by modelling the surface brighrtness distribution of the galaxies and assuming a particular stellar mass-to-light ratio.  \citet{geehan06} find $D/T \approx 0.69$ for M31 and \citet{dehnen98} find $D/T \approx 0.73-0.92$ (the exact value depending on the surface brightness modelling assumptions).  These values fall within the simulated D/T distribution when the dynamical method of C10 is used but they are larger than any of values of D/T estimated using the method of \citet{abadi03} with the angular momentum cut of \citet{zolotov09}.  This begs the question, which is the appropriate comparison to make?  The answer is likely to be neither.  As pointed out recently by \citet{scannapieco10}, there can be large systematic differences in the D/T ratios inferred through kinematic and `morphological' (i.e., based on surface brightness modelling) methods.  In particular, those authors found by analysing simulated galaxies in the same way as observers do (i.e., morphologically, including dust attenuation) they typically recovered much higher D/T ratios than that recovered with the standard dynamical decomposition applied by simulators.  Thus, to make a more meaningful comparison between our simulated galaxies and M31 and the Milky Way would require analysing our simulations in the same way as has been done for these two galaxies.  This is clearly a worthwhile exercise but is beyond the scope of the present study.


Since the present work concerns itself with the properties of the extended stellar distributions of disc galaxies (i.e., the spheroids), it is also important to check whether the simulations yield reasonable satellite luminosity functions and radial distributions, since a significant fraction of the spheroid (by volume at least) is thought to have been assembled through the tidal disruption of massive satellites.  In Fig.~\ref{fig:sat_LF} we compare the mean satellite luminosity functions (LFs) of simulated galaxies at $z=0$ (both for all galaxies in our sample and for a subset of high-mass simulated galaxies) with those of the Milky Way and M31.  For the simulated galaxies the LF is calculated for all satellites within $300$ kpc, similar to the outermost radius adopted for the observational data.  The LFs of the satellite systems of the Milky Way and M31 have a slightly higher normalisation than the mean LF derived from all the simulated galaxies.  However, encouragingly, they are quite compatible with a large subset of massive galaxies in our sample, with masses in the range $1.8\times10^{12} \la M_{200}/M_\odot \la 3.2\times10^{12}$ (128 disc galaxies meet this criteria).  The observationally inferred masses for the MW and M31 are fully consistent with this range (e.g., \citealt{guo10}).  Note that the flattening in the LF of the simulated satellites at $M_V \ga -12.5$ is due to the finite resolution of our simulations (see the Appendix for a convergence study).
In terms of the spatial distribution, \citet{deason11c} examined the cumulative radial distribution of the 10 brightest satellites around Milky Way-mass galaxies in the intermediate-resolution \gimic\ simulations.  They find that the Milky Way and M31 distributions fall within the system-to-system scatter of the simulated galaxies.  Taken together, these results imply that the simulated spheroids should have realistic contributions from the accretion and disruption of satellites.

We now proceed to our analysis of the stellar spheroids of simulated disc galaxies.

\begin{figure*}
\includegraphics[width=\columnwidth]{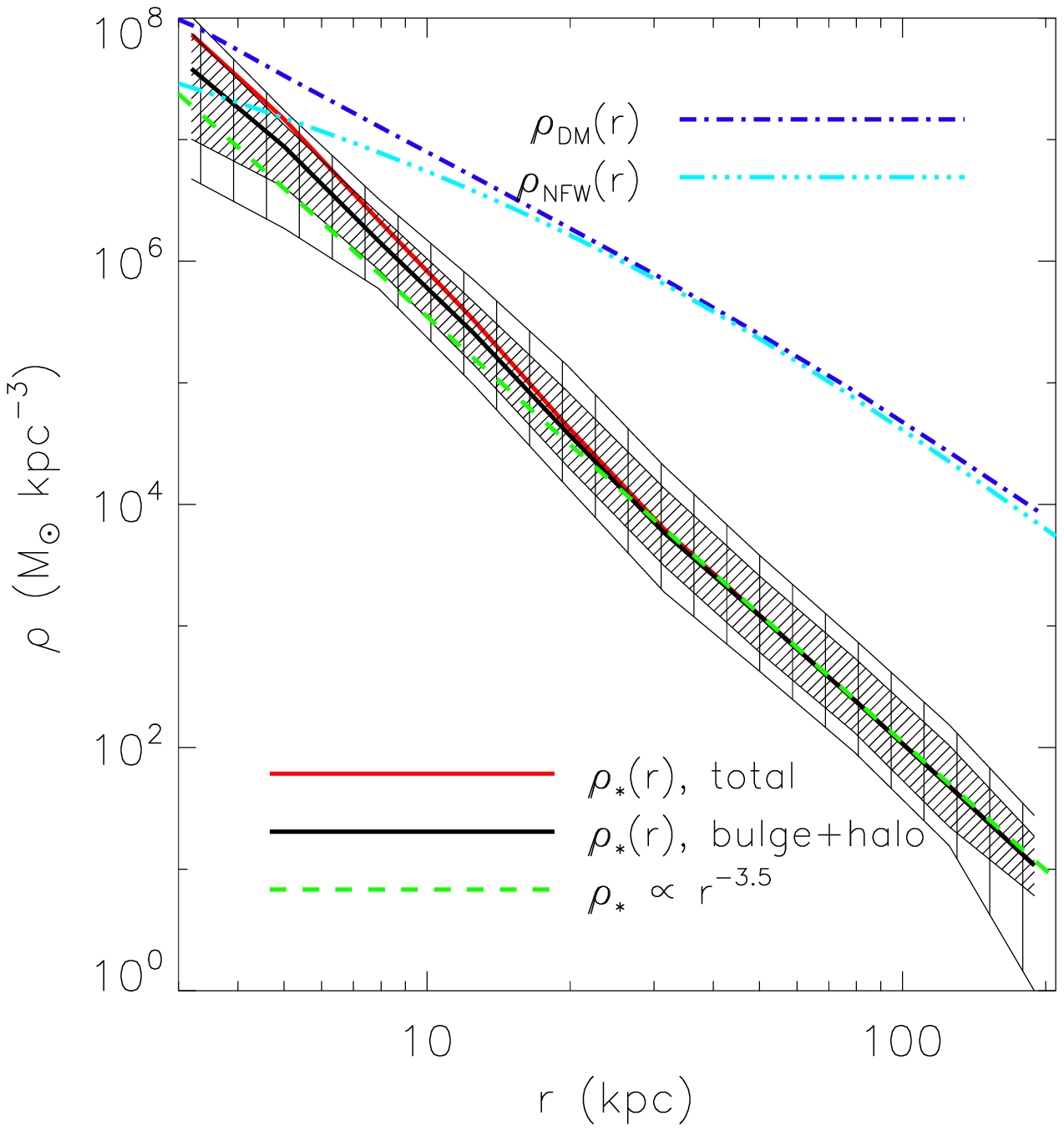} 
\includegraphics[width=\columnwidth]{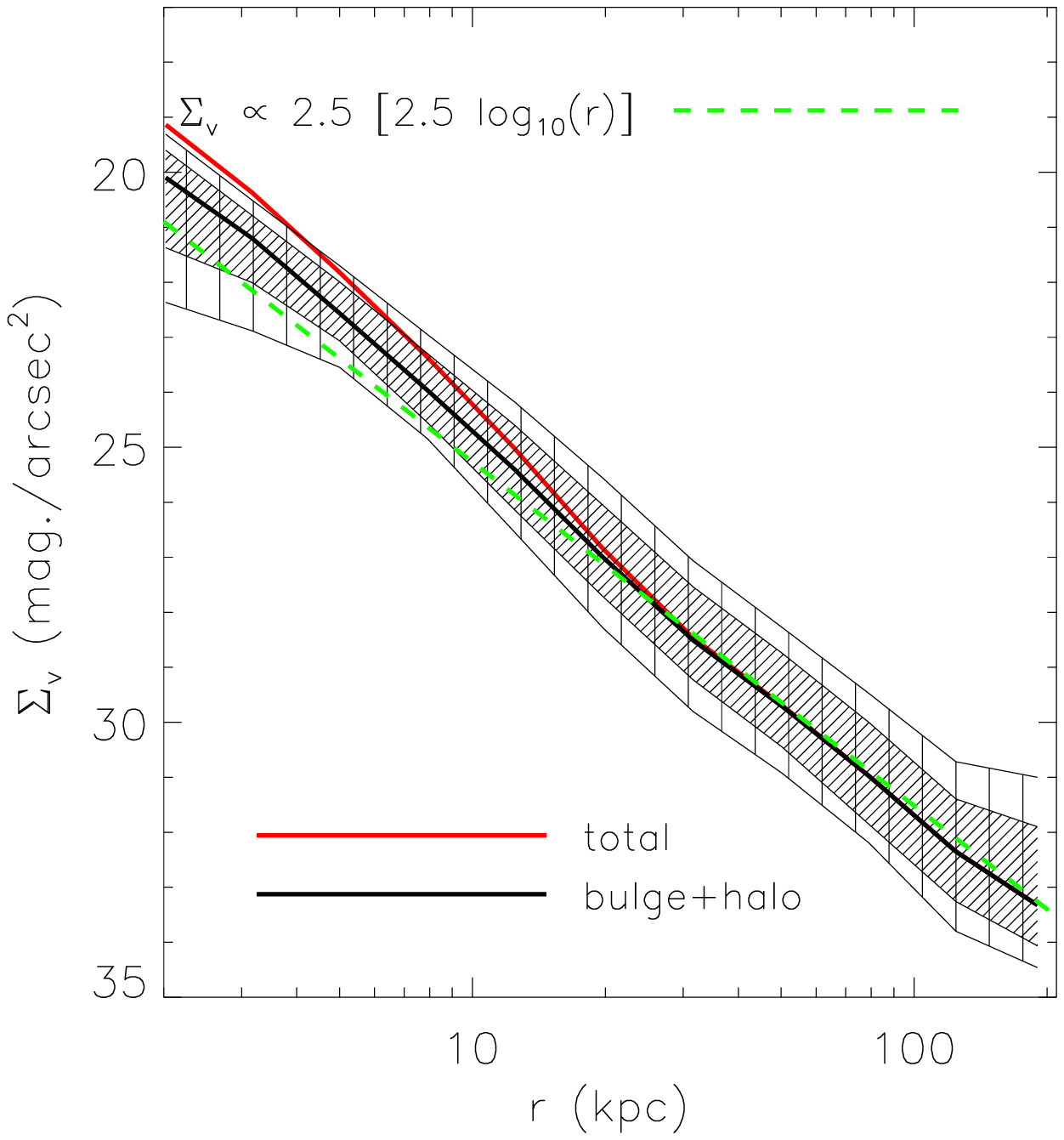}
\caption{\label{fig:rho_profs} {\it Left:} Median spherically-averaged mass density profiles.  The solid red curve is the median stellar mass density profile for all stars, whereas the solid black curve is the median stellar mass density profile for just the spheroidal (bulge+halo) component.  The finely hatched region encloses the $14^{\rm th}$ and $86^{\rm th}$ percentiles of the stellar bulge+halo mass density profile, while the sparsely hatched region encloses the $5^{\rm th}$ and $95^{\rm th}$ percentiles.  The dashed green curve represents a power-law profile with $\rho_{*} \propto r^{-3.5}$ with a normalisation chosen to match the median stellar halo mass density profile at large radii.  The dot-dashed blue curve represents the median dark matter mass density profile, while the dot-dashed cyan curve represents a \citet{navarro96} mass density profile with $M_{200} = 1.3 \times 10^{12} M_\odot$, which corresponds to the median halo mass of our sample of simulated galaxies.  {\it Right:} Median azimuthally-averaged surface brightness profile in the $V$-band for the randomly-oriented simulated galaxies.  The solid red curve is the median surface brightness profile for all stars, whereas the solid black curve is the median profile for the bulge+halo.  The finely (respectively sparsely) hatched region encloses the $14^{\rm th}$ and $86^{\rm th}$ (respectively $5^{\rm th}$ and $95^{\rm th}$) percentiles of the bulge+halo surface brightness profile.  The dashed green curve represents a power-law profile (functional form given in the legend) with a normalisation chosen to match the median surface brightness profile at large radii.  No single power-law can match the simulated bulge+halo profiles over all radii.  At radii of $r \la 30$ kpc, a `bump' is present, signaling the transition from accretion-dominated to in situ-dominated stars.  
}\end{figure*}

\subsection{Structural properties}

Below we present the radial distributions of stellar mass density, $\rho_{*}$, and surface brightness, $\Sigma_V$, as well as the radial profiles of the metallicity [Fe/H] and [$\alpha$/Fe] of our sample of simulated galaxies.

\subsubsection{Stellar mass density and surface brightness profiles}

Fig.~\ref{fig:rho_profs} shows the median spherically-averaged stellar mass density (left panel) and azimuthally-averaged V-band surface brightness (right panel) radial profiles of the bulge+halo component (solid black curves), as well as the 1- and 2-sigma scatter about the median (shaded regions).  The solid red curves show the median profiles for the {\it total} stellar (disc+bulge+halo) components.  The dashed green curve represents a power-law in $\rho_{*}$ with index of $-3.5$ in the left panel, which corresponds to a power-law in surface brightness (in linear units) with index of $-2.5$, assuming the stellar mass-to-light ratio is independent of radius.  Converting to mags. per arcsec$^2$, implies a power-law with index 5.25, which is represented by the dashed green curve in the right panel.

First, extended (out to $r \sim 100$ kpc and beyond) stellar distributions are a ubiquitous feature of the simulated galaxy populations.  The radial distributions of mass density and surface brightness have relatively simple forms, but as a comparison of the solid black and dashed green curves demonstrates, no single power-law distribution can match the profiles over all radii.  A power-law with index of $-3.5$ (in $\rho_*$) provides an excellent fit to the outer regions, $r \ga 30$ kpc, of the median bulge+halo profile, but the same power-law under-predicts the mass density/surface brightness at smaller radii.  As we will discuss in Section \ref{sec:origin}, the change in the nature of the radial distributions at radii of $r \approx 30$ kpc is driven by a change in the relative contributions of stars deposited by tidally disrupted satellites and stars that formed in situ.  

We can quantify the system-to-system scatter in the slope of the profile at large radii by fitting power-laws to the mass density distributions of all $\approx 400$ disc galaxies in our sample.  We find a median slope of $-3.51$ with $-3.05$, $-3.20$, $-3.89$, and $-4.16$ as the 5th, 14th, 86th, and 95th percentiles of the distribution of slopes, respectively.  Although it usually does not fit the profiles as well as a power-law, we have also fit a de Vaucouleurs profile to the outer regions (as done, e.g., by \citealt{ibata07} for the outer regions of M31).  In this case we find a median effective radius, $r_e$, of $17.80$ kpc with $6.71$, $10.05$, $34.74$, and $54.98$ kpc as the 5th, 14th, 86th, and 95th percentiles, respectively.

In the inner regions, $r \la 30$ kpc, we find that a de Vaucouleurs profile usually fits the mass density distribution very well.  Here we find a median $r_e$ of $1.83$ kpc with $0.73$, $0.95$, $4.90$, and $7.54$ kpc as the 5th, 14th, 86th, and 95th percentiles, respectively.  A power-law fit to $r \la 30$ kpc yields a median slope of $-2.98$ and $-2.47$, $-2.58$, $-3.30$, and $-3.36$ as the 5th, 14th, 86th, and 95th percentiles, respectively.  This is shallower than in the outer regions, but note that the mass density profile has a higher normalisation at small radii, so that a power-law with slope $-3.5$ extrapolated to the inner regions underpredicts the mass density there.

In general, our results confirm those of several previous theoretical studies on the shape of the mass density profile at larger radii.  For example, \citet{bullock05}, who used a hybrid semi-analytic plus idealised N-body approach to study the formation of the stellar halo, find that a power-law with index of $-3.5$ describes the shape of their haloes very well at radii of $\approx50-100$ kpc.   More recently, \citet{cooper10} applied a similar approach to the very high-resolution Aquarius cosmological N-body simulations.  These authors did not report on a power-law fit to their simulated profiles at large radii, but an examination of their Fig.\ 4 suggests a slightly steeper logarithmic slope of $\approx -4$.   The shape of the bulge+halo component at large radii is also similar to that found by \citet{abadi06}, who performed cosmological SPH re-simulations of a small number of Milky Way-mass systems.  In particular, these authors found logarithmic slopes of $-2.9$ at $100$~kpc and $-3.5$ at the virial radius of their simulated galaxies.

While there is generally good agreement at large radii between our results and those of previous theoretical studies, some differences become apparent at small radii.  In particular, \citet{bullock05} find that their profiles become increasingly flat at small radii, so that the logarithmic slope approaches $-1$ for $r \sim 10$ kpc, whereas our spheroid component shows no evidence of flattening at small radii - on the contrary, the profile actually steepens somewhat at radii of $r \approx 30$ kpc before leveling off to a slope of $\approx -3$ at $r \la 20$ kpc.  This dramatic flattening in the \citet{bullock05} model is almost certainly in part due to the neglect of in situ star formation in their hybrid N-body model, but another potentially important factor is the treatment of the galaxy potential well, which does not feel the back-reaction due to the orbiting satellites in their idealised N-body simulations.  For example, \citet{cooper10} (see also \citealt{diemand05}), who applied a similar hybrid method to cosmological N-body simulations with `live' galaxy potential wells, find a relatively steep distribution at small radii, with a logarithmic slope of $\approx -3$.  This suggests that the treatment of the potential well is indeed relevant.  Consistent with this hypothesis is that when we fit a power-law to the density profile {\it due to accreted stars only} at radii of $r \le 30$ kpc, we still find a steep logarithmic slope of $\approx -2.9$ (i.e., similar to that found by \citealt{cooper10}).  The fact that our {\it total} (accreted+in situ) profiles show evidence for an increase in normalisation at small radii (at $r \approx 30$ kpc), whereas \citet{cooper10} find, if anything, a slight flattening, suggests that in situ star formation also plays an important role at small radii.  We demonstrate that this is indeed the case in Section \ref{sec:origin}.

\begin{figure}
\includegraphics[width=\columnwidth]{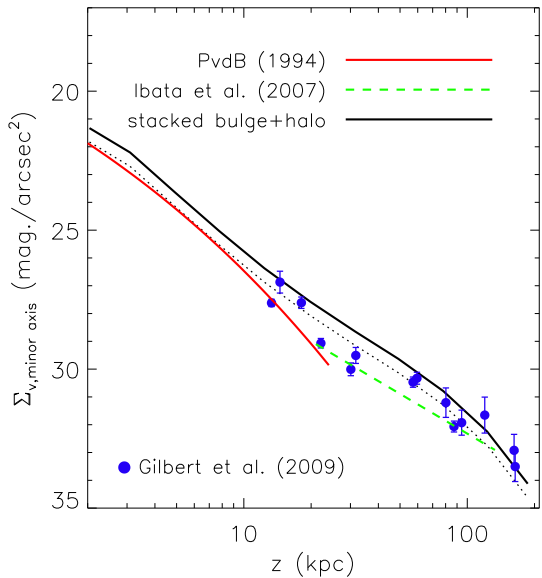}
\caption{\label{fig:sig_profs} Comparison of the minor axis V-band surface brightness profile for the simulated galaxies with that of M31.  The solid black curve represents the stacked profile derived from all 412 simulated disc galaxies. The dotted black curve is the stacked simulated profile shifted down by 0.5 mags arcsec$^{-2}$.  The solid red curve represents the de Vaucouleurs profile fit to the minor axis profile of the M31 ``bulge'' derived by (\citealt{pritchet94}, PvdB).  The dashed green line represents the power-law fit to the outer halo of M31 by \citet{ibata07}.  The solid blue circles represent the measurements of \citet{gilbert09a}.  The shape of the stacked profile is in very good agreement with that derived for M31 over a wide range of radii.
}\end{figure}

It is interesting to compare the mass density/surface brightness profiles of the simulated galaxies with those inferred for local galaxies.  We first consider M31, which may represent the best test of the models, as its relatively close location and our external view point allow observations of low surface brightness levels down to $ \sim 32$ mag arcsec$^{-2}$ and hence enable the tracing of the radial profiles out to large projected radii.  

In Fig.~\ref{fig:sig_profs} we compare the V-band surface brightness profiles of the simulated galaxies to that derived for M31.  To date the surface brightness profile of M31 has mainly be derived along (or close to) the minor axis.  For the purposes of comparison we therefore also measure the minor axis profile of our simulated galaxies.   This is done by orienting the simulated galaxies to an edge-on configuration (with $x-y$ plane being parallel to the plane of the disc) and measuring the surface brightness as a function of the distance $z$ from the disc midplane within a cylindrical distance ($R = \sqrt{x^2+y^2}$) of 5 kpc from the system center.  Unfortunately, the mass resolution of our simulations prevents us from measuring the minor axis profile on a system-by-system basis\footnote{This was possible for the azimuthally-averaged profiles in Fig.~\ref{fig:rho_profs} because of the much larger surface area used for measuring the profiles.}, so we have stacked all 412 edge-on galaxies to derive a mean minor-axis surface brightness profile.  This is represented by the solid black curve in Fig.~\ref{fig:sig_profs}.  For comparison, we show the minor axis surface brightness measurements of \citet{pritchet94}, \citet{ibata07}, and \citet{gilbert09a}.

Quite remarkably, the shape of the stacked simulated profile matches that of the observed surface brightness profile of M31 over almost two decades in radius (from $r \approx 2$ kpc to beyond 100 kpc), including the change in slope that occurs at radii of $20-30$ kpc.  

Note that it is the {\it shape} of the profile, rather than its normalisation, that is the more critical test of the models, since the normalisation of the observed profile requires making an uncertain conversion between the tracer population (Red Giant star counts in this case) to the total surface brightness.  Typically this is done by assuming that the spheroid has an identical stellar luminosity function to that of local dwarfs or globular clusters and therefore one can use the relationship between integrated light and the same tracers in these local systems to re-normalise the halo.  Of course this procedure will only be accurate if the stellar halo indeed has a similar stellar luminosity function to the local dwarfs or globular clusters.  For the data plotted in Fig.~\ref{fig:sig_profs}, \citet{pritchet94} used the observations of the Milky Way globular cluster 47 Tuc for the conversion, while \citet{ibata07} and \citet{gilbert09a} shifted their data vertically to map onto the profile \citet{pritchet94} at radii of $\approx 10-30$ kpc.  In spite of the potential for important systematics in normalisation of the observed profile, the stacked profile agrees with the data to an accuracy of $\approx 0.5$ mags. arcsec$^{-2}$.  This is actually better than one might have naively expected, especially given that the system-to-system scatter in the simulated profiles is $\approx 3$ mags. arcsec$^{-2}$ (see Fig.~\ref{fig:rho_profs}).

We turn now to the Milky Way.  Bearing in mind that observations of the extended stellar distribution of the Milky Way are typically limited to $r \la 40$ kpc, a wide variety of studies have found logarithmic slopes for the density profile of the stellar halo ranging from from $-2.5$ to $-3.75$, depending on the data and methodology that were used \citep{harris76,zinn85,saha85,preston91,wetterer96,ivezic00,morrison00,yanny00,chiba00,chen01,siegel02,vivas06,bell08,juric08,smith09,sesar11,deason11b}.  This is quite compatible with results of the cosmological hydrodynamical simulations.  Interestingly, some of the latest observations of the Milky Way halo, which are based on large samples of stars of the SDSS, find evidence for a change in the slope density profile at radii of around $\approx 30$~kpc \citep{carollo07,juric08,sesar11,deason11b}, which is remarkably consistent with our simulations.  The same studies also find strong evidence that the inner halo has an oblate distribution (rather than spherical), with a minor/major axis ratio of $0.5-0.7$.  We present a study of the shape and kinematics of the simulated galaxy haloes in McCarthy et al.\ (in prep.).  However, a visual inspection of the simulated azimuthally-averaged and minor-axis profiles in Fig.~\ref{fig:rho_profs} and \ref{fig:sig_profs} already reveals that our spheroids are indeed flattened within 30 kpc or so.

Observations of stellar haloes of other (more distant) external galaxies are more difficult. Currently, only a few stellar haloes of individual galaxies have been detected, and only down to levels of $\sim 28-29$ mag arcsec$^{-2}$ \citep{dejong07}, corresponding to the inner regions.   One can go deeper by stacking distant galaxies.   For example, \citet{zibetti04} stacked more than 1000 stellar haloes of edge-on spiral galaxies observed in the SDSS and found that, out to about $25$~kpc ($\Sigma_V \sim 31$ mags. arcsec$^{-2}$), the average density profile can be fitted relatively well with a power-law $\rho_{*} \sim r^{-3}$, very similar to what the simulations predict. 

\begin{figure*}
\includegraphics[width=\columnwidth]{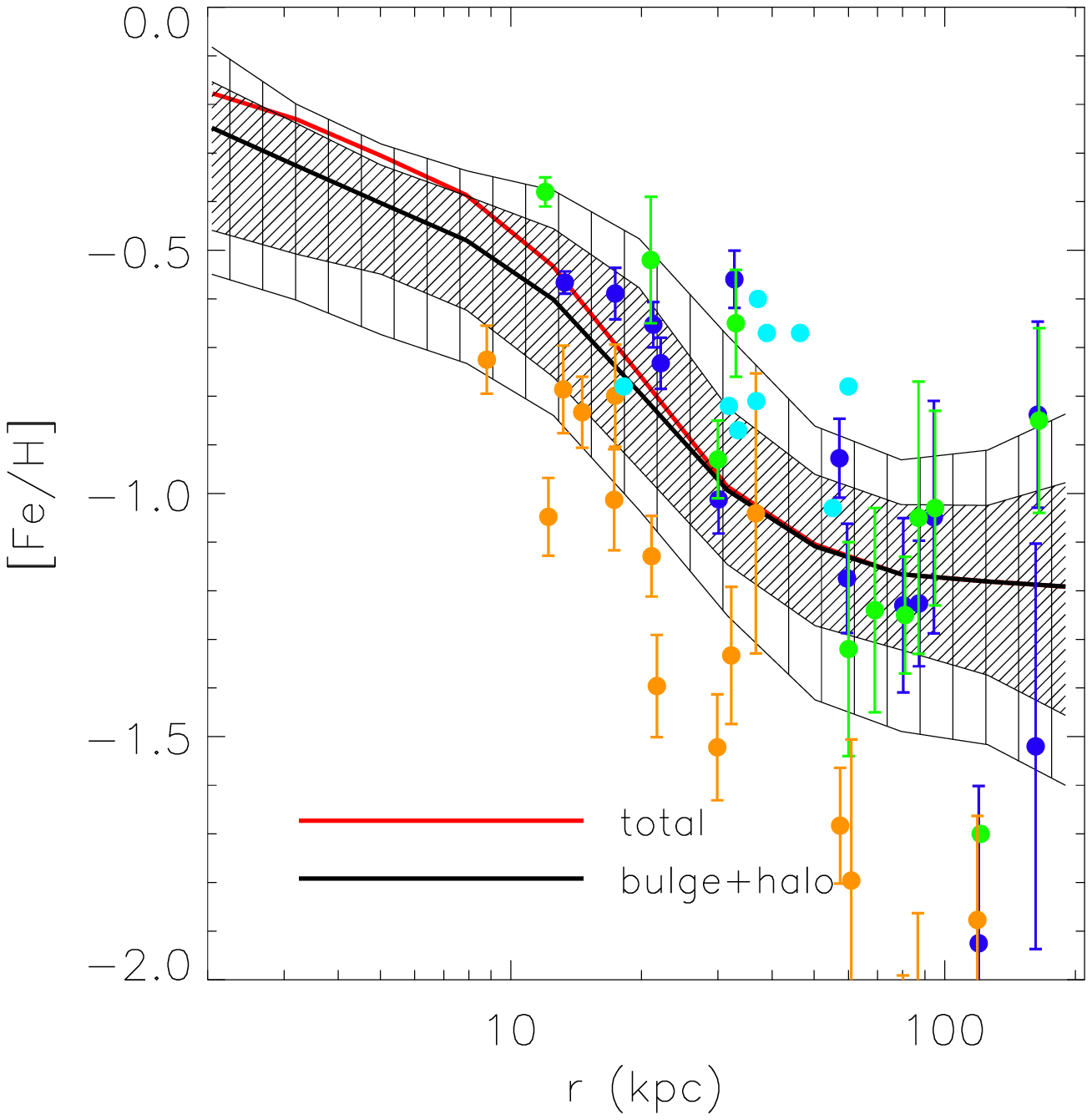}
\includegraphics[width=\columnwidth]{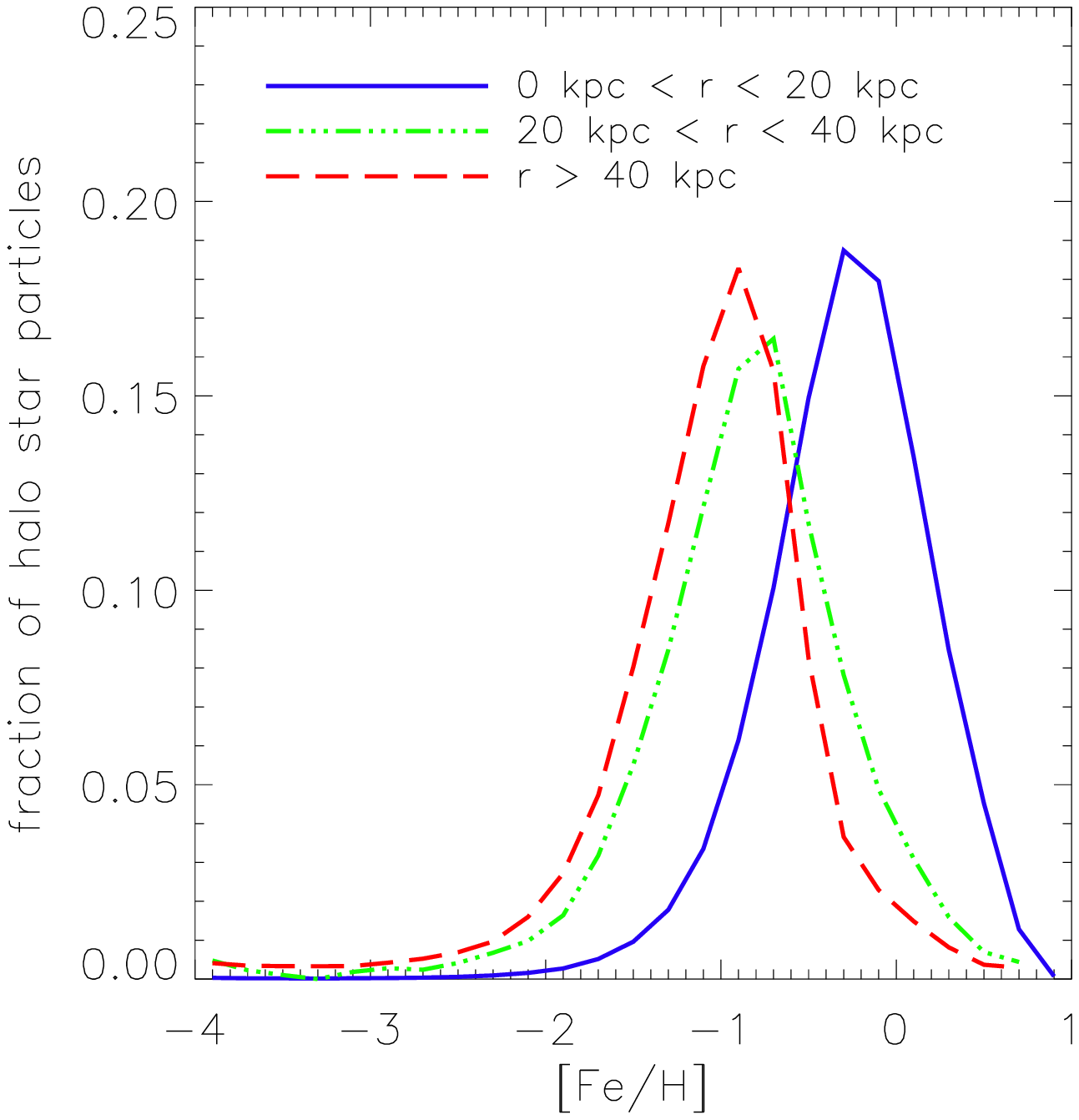}
\caption{\label{fig:iron_profs} {\it Left}: Median spherically-averaged stellar metallicity profiles.  The solid red curve is the median stellar metallicity profile for all stars, whereas the solid black curve is the median stellar metallicity profile for just the bulge+halo component.  The finely (sparsely) hatched region encloses the $14^{\rm th}$ and $86^{\rm th}$ ($5^{\rm th}$ and $95^{\rm th}$) percentiles of the stellar halo metallicity profile.  The solid blue, green, cyan, and orange circles represent M31 metallicity measurements of \citet{gilbert09a}, \citet{kalirai06}, \citet{richardson09}, and \citet{koch08}, respectively.  The observational data have been scaled to the same set of solar abundances.
{\it Right:} Median metallicity distribution function (MDF) in three radial bins. Significant negative metallicity gradients are a ubiquitous feature of the simulated galaxy populations.}
\end{figure*}

With our large sample of simulated galaxies ($\approx 400$), we are in an excellent position to characterise the system-to-system {\it scatter in the normalisation} of the $\rho_*$ and $\Sigma_V$ profiles, which is another potential test of the simulations. The solid black curves represent the median profiles of the bulge+halo components only.  The finely-shaded regions enclose the $14^{\rm th}$ and $86^{\rm th}$ percentiles (i.e., 1$\sigma$) and the sparsely-shaded regions enclose the $5^{\rm th}$ and $95^{\rm th}$ percentiles (2$\sigma$) of the stellar bulge+halo profiles.  The $2\sigma$ scatter in $\rho_*$ is approximately an order of magnitude over the range of $10 - 100$kpc, corresponding to a scatter in $\Sigma_V$ of $\sim 3$ mags./arcsec$^2$ over the same range. [In the Appendix, we test the numerical convergence of the stellar mass density profiles.  We demonstrate there that our results (both median and scatter) are quite robust to a factor of eight increase in the mass resolution.]

It is also worth commenting briefly on the structure of the dark matter halo.  In addition to the lines corresponding to the stellar components described above, the left panel of Fig.~\ref{fig:rho_profs} also shows the median dark matter density distribution (as the blue curve) and an analytic NFW \citep{navarro96} profile (the cyan curve). The analytic halo has a mass $M_{200} = 1.4 \times 10^{12} M_\odot$ (which corresponds to the median halo mass of our sample of simulated galaxies) and a concentration parameter of $c_{200} = 9.4$, which places it on the $M_{200} - c_{200}$ relation measured by \citet{gao08} for the Millennium Simulation.  The dot-dashed cyan curve {\it is not a fit} to the dot-dashed blue curve.

A comparison of the dot-dashed blue curve with the solid red curve demonstrates that dark matter dominates over the stellar contribution down to very small radii (the two become comparable at a few kpc).  Comparison of the dot-dashed blue and cyan curves shows that the NFW analytic form provides an excellent description of the dark matter density profile beyond about $r \approx 10$ kpc (i.e., $\approx 0.05 r_{200}$).   Within this radius, the dark matter profile steepens, presumably as a result of contraction of the dark matter halo due to the condensation of baryons (e.g., \citealt{blumenthal86,gnedin04,duffy10}).  

Given that the NFW profile drops off as $\sim r^{-2}$ at intermediate radii and as $\sim r^{-3}$ at large radii, this implies that the stellar mass density profile drops off significantly faster than that of the dark matter, as is clearly visible in the left panel of Fig.~\ref{fig:rho_profs}.  This is also consistent with the findings of previous theoretical studies of the stellar halo (e.g. \citealt{bullock05,abadi06}).  The difference in the stellar and dark matter distributions likely stems from the fact that the stars in infalling satellites are more tightly bound, and thus less susceptible to tidal disruption, than their dark matter haloes.  In addition, the most massive satellites (which also have the highest stellar mass fractions) will spiral more quickly into the center due to dynamical friction, which scales as mass squared.  Finally, in situ star formation will occur preferentially in the central regions (due to the higher gas densities there).  These effects would produce a more centrally-concentrated stellar distribution, as seen in the simulations as well as in the observations.

\subsubsection{ {\rm \FeH} and {\rm \alphaFe} radial profiles}

The radial distribution of stellar metallicity is a particularly interesting test of the simulations, since previous simulations/models in the $\Lambda$CDM context have not been able to reproduce the significant gradients seen in M31 and the Milky Way.  It is not presently clear why this is the case.  It may signal a fundamental problem with galaxy formation in the $\Lambda$CDM context or that previous models of stellar haloes neglected important physics, such as in situ star formation or a sufficiently realistic implementation of chemodynamics.  

In the left panel of Fig.~\ref{fig:iron_profs} we show the median spherically-averaged metallicity profile and system-to-system scatter about the median.  The metallicity, [Fe/H], of each star particle\footnote{For consistency with the way the chemodynamics is done in the simulations, we used so-called `smoothed metallicities' rather than `particle metallicities'.  See \citet{wiersma09b} for discussion.  Switching to `particle metallicities' has the effect of shifting the profiles in the left panel of Fig.~\ref{fig:iron_profs} down by $\approx 0.2$ dex, which, as discussed later in the paper, is smaller than the uncertainties in the adopted empirical nucleosynthesis yields and Type Ia SNe rates.} is computed by taking the logarithm of the ratio of the iron-to-hydrogen mass fractions of the star particle and subtracting the logarithm of the iron-to-hydrogen mass fraction ratio of the Sun (which we assume is 0.00156, consistent with the recent measurements and 3D modelling of \citealt{asplund05}).  A spherically-averaged metallicity profile is computed for each galaxy by computing the mean of metallicity (i.e., $<$[Fe/H]$>$) of star particles in radial bins.  The median profiles plotted in Fig.~\ref{fig:iron_profs} are then computed by taking the median of all of individual metallicity profiles in radial bins.

Interestingly, the stellar metallicity distribution shows a prominent negative gradient over all radii, with the steepest decline exhibited over the range $20$ kpc $\la r \la 40$ kpc.  This is independent of whether the disc component is included or not.  The solid black curve and shaded regions indicate that {\it negative metallicity gradients are a ubiquitous feature of the simulated galaxy populations}.  (This statement is further confirmed in the middle panel Fig.~\ref{fig:frac_insitu} in Section \ref{sec:origin}, which shows system-to-system scatter in the difference in the mean metallicity within and beyond 30~kpc.)

\begin{figure*}
\includegraphics[width=\columnwidth]{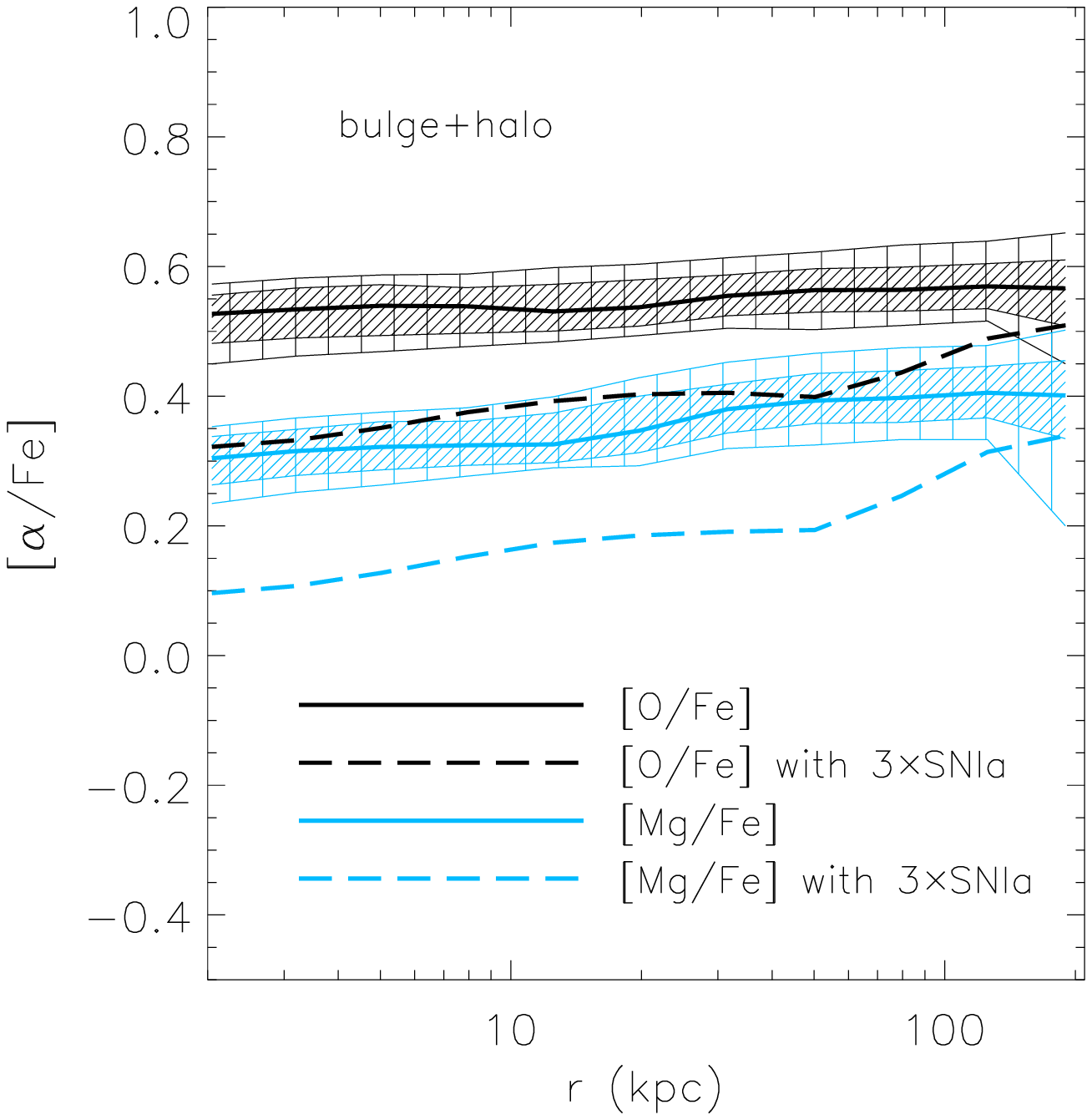}
\includegraphics[width=\columnwidth]{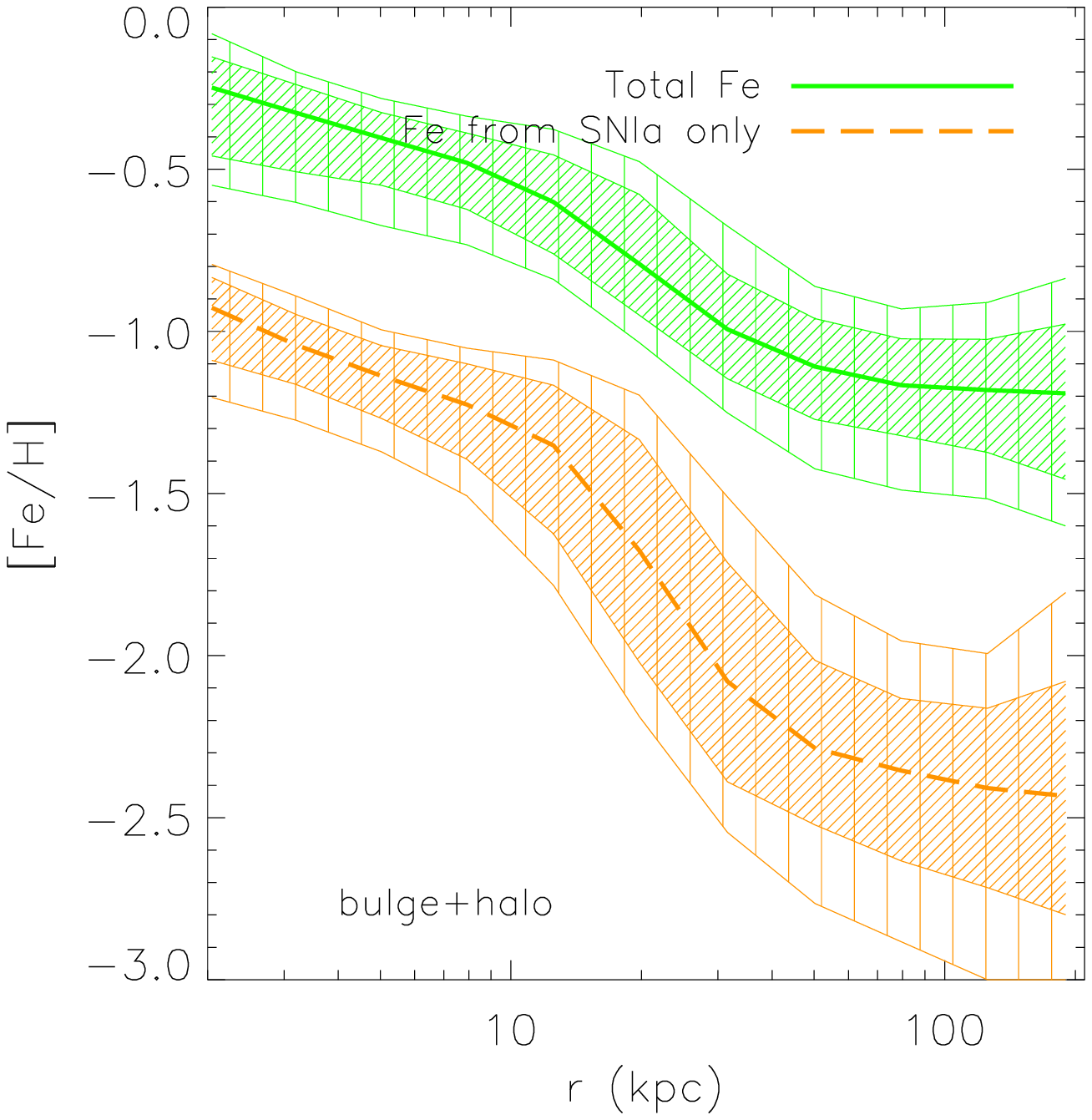}
\caption{\label{fig:alphafe_profs} {\it Left}: Median spherically-averaged stellar [$\alpha$/Fe] profiles.  The solid black (light blue) curve is the median [O/Fe] ([Mg/Fe]) profile for the bulge+halo component.  The finely (sparsely) hatched region encloses the $14^{\rm th}$ and $86^{\rm th}$ ($5^{\rm th}$ and $95^{\rm th}$) percentiles.  The dashed black (light blue) curve shows the effect on the median [O/Fe] ([Mg/Fe]) profiles if the rate of Type Ia SNe were to be increased by a factor of 3 in the simulations.  {\it Right}: Median spherically-averaged stellar [Fe/H] profiles for the spheroidal component using the total mass of iron (green) and just the iron produced by Type Ia supernovae (orange).  The finely (sparsely) hatched regions enclose the $14^{\rm th}$ and $86^{\rm th}$ ($5^{\rm th}$ and $95^{\rm th}$) percentiles.  There is no significant radial variation in the simulated [$\alpha$/Fe] profiles, owing to the dominance of metal production by Type II SNe at all radii.}
\end{figure*}

The solid blue, green, cyan, and orange circles in Fig.~\ref{fig:iron_profs} represent metallicity measurements of M31 from \citet{gilbert09a}, \citet{kalirai06}, \citet{richardson09}, and \citet{koch08}, respectively\footnote{Where appropriate we have adjusted the observational metallicity measurements (by applying a constant offset) to account for differences in the assumed solar abundances in the studies of \citet{gilbert09a}, \citet{kalirai06}, \citet{richardson09}, and \citet{koch08} from the solar abundances of \citet{asplund05}, which we use to normalise the simulated profiles.}.  The simulated profiles have a remarkably similar shape to that derived by \citet{gilbert09a} and \citet{kalirai06}, which shows a drop of $\approx 0.7-0.8$ dex over the wide radial range of $\approx10-150$ kpc.  The simulated profiles are also consistent with the measurements of \citet{richardson09} (see also \citealt{chapman06}), which probe a much smaller dynamic range.  The profile derived by \citet{koch08}, however, is somewhat steeper than the predicted profiles, particularly at very large radii.  It is worth noting that 
\citet{gilbert09a}, \citet{kalirai06}, and \citet{richardson09} all employed similar methods in deriving the metallicity (in particular, they used spectroscopically-calibrated photometric methods), whereas \citet{koch08} used a new spectroscopic method that derives abundances from the calcium triplet (CaT) part of the spectrum.  It is not presently clear what the origin is of the difference between the different data sets.

It is also encouraging that the normalisation of the simulation metallicity profiles agrees as well as it does (to within 0.2-0.3 dex accuracy) with the observations of M31.  The simulations adopt empirical nucleosynthesis yields and Type Ia SNe rates, both of which are uncertain by a factor of a few (see \citealt{wiersma09b}), implying that there is freedom to vertically shift the simulated profiles up or down by a few tenths of a dex.  Another potentially relevant factor is the selection of the observational fields.  Often they are selected to have a relatively high surface brightness.  \citet{font08} and \citet{gilbert09a} have shown that surface brightness tends to be positively correlated with metallicity, so it may well be the case that the current measurements are biased somewhat high.  Large panoramic observations of M31 (such as PAndAS) will soon shed light on whether this bias is important.

For the Milky Way, metallicity measurements out to large galactocentric radii are much more challenging.  However, recent results based on SDSS observations suggest suggest a similar drop of $\approx 0.7-0.8$ dex from the Solar neighbourhood out to $\approx 30-40$ kpc (\citealt{carollo07,ivezic08,carollo10,dejong10}, but see \citealt{sesar11}).

As alluded to above, most previous theoretical studies, particularly those based on accretion-only hybrid models (e.g., \citealt{font06a,delucia08,cooper10}), did not find significant gradients.  To our knowledge, previous studies based on hydrodynamic re-simulations of disc galaxies (e.g., \citealt{abadi06,zolotov10}) did not examine the radial distribution of the stellar metallicity\footnote{But note that hydrodynamic re-simulations have been used to study the radial distribution of metallicity in {\it early-type} galaxies (e.g., \citealt{kobayashi04,kobayashi05,gibson07}) and significant gradients were shown to be common in these systems as well.}.  As we demonstrate in Section \ref{sec:origin}, the change in slope at $r \sim 30$ kpc marks the transition from accretion ($r \ga 30$ kpc) to in situ ($r \la 30$ kpc) stellar halo formation.

While the spherically-averaged profiles of the simulated galaxies display prominent gradients, at any given radius there is a wide range of stellar metallicities. To illustrate the spread in metallicities as a function of radius, we show the median stellar metallicity distribution function (MDF) for three radial bins in the right panel of Fig.~\ref{fig:iron_profs}.  For all three bins the MDF is similar in shape (in particular, fitting a Gaussian distribution yields similar widths of $\sigma \approx 0.5$~dex), but the peak shifts to lower metallicities with increasing radius, which is what gives rise (mathematically at least) to the metallicity gradients in the left panel of Fig.~\ref{fig:iron_profs}.  A Gaussian provides a reasonable match to the MDFs near the peak and for higher metallicities, but it systematically under-predicts the fraction of halo stars with low metallicities.  The MDFs in the right panel of Fig.~\ref{fig:iron_profs} are qualitatively similar to that found for the M31 stellar halo (Gilbert et al.\ in prep) and other nearby galaxies \citep{mouhcine05b}.

Since the simulations include enrichment from both Type II and Type Ia supernovae, as well as stellar mass loss (i.e., AGB stars), we can also make predictions for the [$\alpha$/Fe] radial profiles, which should be measurable in galaxies like M31 in the near future.  The so-called $\alpha$ elements (such as oxygen, magnesium, and silicon) provide an additional constraint, as they are produced primarily in massive stars.  In the left panel of Fig.~\ref{fig:alphafe_profs} we show the median spherically-averaged [$\alpha$/Fe] profile and scatter, where we have used both oxygen and magnesium to represent $\alpha$.  We have normalised the profiles using the oxygen-to-iron and magnesium-to-iron mass fractions of the Sun (5.2887 and 0.5355, respectively, \citealt{asplund05}).  Interestingly, while the metallicity (i.e. [Fe/H]) profile shows a strong radial dependence (Fig.~\ref{fig:iron_profs}), the [$\alpha$/Fe] distribution does not: [$\alpha$/Fe] rises by only $\approx 0.1$ dex from the center to the outskirts.  This implies that the physical mechanism responsible for producing the iron does not change appreciably with galactocentric distance.

It is interesting to elucidate the lack of a strong radial dependence in the [$\alpha$/Fe] profiles presented in the left panel of Fig.~\ref{fig:alphafe_profs}.  The \gimic\ simulations track not only the total iron mass fraction, but also the iron mass fraction produced by Type Ia supernovae only.  This therefore allows us to determine, for any given star particle, the individual contribution of Type Ia to the iron mass fraction.  In the right panel of Fig.~\ref{fig:alphafe_profs} we show the [Fe/H] profiles for the total iron mass fraction and for the iron produced solely by Type Ia supernovae.  As it can be clearly seen, Type Ia supernovae make only a minor contribution to the metallicity of the stellar halo.  Type II supernovae are therefore the dominant producers of metals at all radii.  The very mild gradient in [$\alpha$/Fe] is due to the relatively larger importance (though still small in an absolute sense) of Type Ia for stars located in the inner regions.

It is important to acknowledge potential caveats in the above prediction.  Besides the nucleosynthesis yields, the rate of Type Ia supernovae is very uncertain.  Rather than try to calculate the rate of Type Ia supernovae using a physical model, the simulations adopt empirical Type Ia rates, which are still uncertain by a factor of a few (see Fig. A6 of \citealt{wiersma09b}).  If the Type Ia rate is actually a factor of a few higher than current estimates imply, this would have the effect of producing a steeper positive [$\alpha$/Fe] gradient (see the dashed curves in the left panel of Fig.~\ref{fig:alphafe_profs}, which show the effect of boosting the number of Type Ia SNe by a factor of 3), since the stars in the central regions were formed more recently (see Section \ref{sec:origin}) and therefore the gas from which they formed had more time to be polluted by Type Ia supernovae.

\section{Physical origin of the radial density and metallicity profiles of the stellar haloes}
\label{sec:origin}

In the previous section we showed that the \gimic\ simulations produce disc galaxies with stellar haloes that bear a strong resemblance to those observed around nearby galaxies, including M31 and our own Milky Way.  In the present section we use the simulations to trace back the physical origin of the stellar haloes.  We first quantify the relative importance of accretion vs.\ in situ star formation for the formation of the stellar halo.  We show that the break in the mass density and surface brightness profiles near $r \sim 30$ kpc and the sharp rise in the stellar halo metallicity profile at similar radii are all caused by the transition from accretion-dominated to in situ-dominated star formation (Section \ref{sec:formation_mode}).  We also explore the time of star formation and assembly of the stellar halo (Section \ref{sec:assembly}) and the degree of radial migration of the in situ component (Section \ref{sec:migration}).

\subsection{Distinguishing accretion and in situ formation}

To assess the relative importance of accretion vs.\ in situ star formation we must first define what we mean by these terms.  We define in situ star formation as star formation that takes place in the most massive subhalo of the most massive progenitor (MMP) FoF group of the present-day system.  Stars are said to be `accreted' if they formed either in another FoF group or in a satellite galaxy (i.e., not the dominant subhalo) of the MMP.  In practice, only a small proportion (2\%) of the stellar halo is from stars that formed in satellites of the MMP and were then stripped; most of the accreted stars were formed in other FoF groups (i.e., other haloes) before they became satellites of the MMP.  

The procedure by which we identify the MMP and determine whether a given star particle was accreted or formed in situ is as follows. For a given Milky Way-mass system at $z=0$ we select all of the dark matter particles within $r_{200}$ belonging to the dominant subhalo.  We use the unique IDs assigned to these particles to identify the FoF group in previous snapshots that contains the largest number of these particles.  This FoF group is said to be the MMP and the dominant subhalo of that FoF group is assumed to be the most massive progenitor of the dominant subhalo of the system at $z=0$.
Each of the star particles that comprise the $z=0$ stellar halo is tracked back in time to the FoF group and subhalo it belonged to, if any, when it formed from a gas particle.  
If, at the time of star formation, the star particle is a member of the dominant subhalo of the MMP, the star particle is said to have formed in situ.  If it formed in another FoF group or in a non-dominant subhalo of the MMP, then it is said to have been accreted\footnote{We have also allowed for the possibility that the star particle may have formed outside of bound substructures but find that this mode of formation is rare, accounting for 2\% of the stellar halo by mass.}.

The number of snapshots output for each of the \gimic\ volumes is approximately 60, and they span the redshift range $z=20$ to $z=0$.  The time of star formation of each star particle will, in general, not correspond to the redshift at which the snapshot data was written.  We therefore must use the snapshot that is closest in time to identify where the star particle was when it formed.  This could, in certain circumstances, lead to a misidentification of the location of star formation (i.e., to which subhalo it belonged when it formed).  However, we have explicitly checked that this is not an important issue by re-running our tracing code using only every second snapshot.  We find that the in situ mass fraction of our simulated galaxies calculated using only every second snapshot agrees with that inferred using all the snapshots to an accuracy of 1.6\% on average.
 We therefore conclude that the redshift sampling of our snapshot data is sufficient to deduce whether a given particle formed in situ or was accreted.

We trace each of the $z=0$ stellar halo particles back in time as far as $z=10$ to determine whether they formed in situ or were accreted.  We ignore the very small fraction (0.01\%) of halo stars that formed before $z=10$. 

\begin{figure}
\includegraphics[width=\columnwidth]{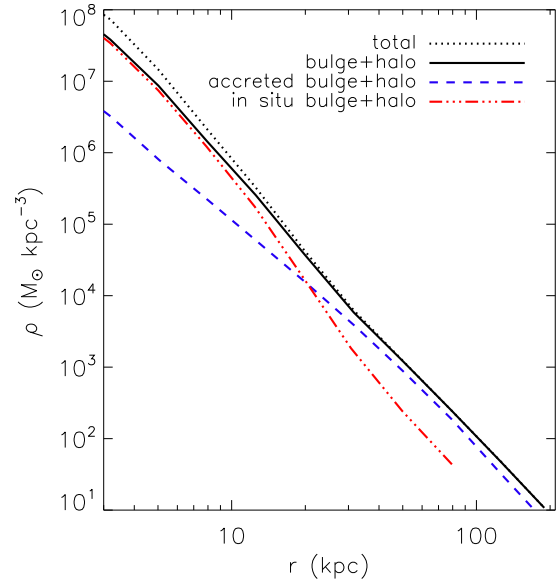}
\caption{\label{fig:rho_profile_state} Median spherically-averaged stellar mass density profiles.  The dotted and solid black curves represent all stars and stars in the spheroid only, respectively.  The dotted black curve represents all stars (disc+spheroid). The dashed blue curve represents accreted halo stars (i.e., stars that formed outside the most massive progenitor).  The dot-dashed red curve represents spheroid stars that formed in situ.  In situ star formation contributes significantly to (dominates) the stellar halo within a radius of 30 (20) kpc.
} 
\end{figure}

\begin{figure}
\includegraphics[width=\columnwidth]{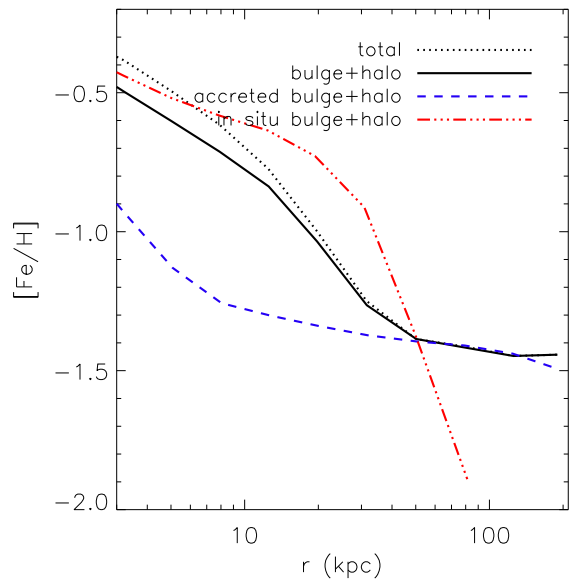}
\caption{\label{fig:FeH_profile_state} Median spherically-averaged metallicity profiles.  The curves have the same meaning as in Fig.~\ref{fig:rho_profile_state}.  For $r < 50$~kpc, stars formed in situ have higher metallicity and there dominance of the stellar halo by mass at small radii is what produces the strong radial variation in [Fe/H].
} 
\end{figure}

\subsection{Stellar mass density and metallicity profiles by formation mode}
\label{sec:formation_mode}

In Fig.~\ref{fig:rho_profile_state} we show the median spherically-averaged mass density profile of systems split according to formation mode (i.e., in situ vs.\ accreted).  A comparison of the dot-dashed red and dashed blue curves demonstrates that in situ star formation contributes significantly to (dominates) the stellar halo within a radius of 30 (20)~kpc.  Beyond 30~kpc the majority of the stars were formed in satellites prior to accretion.  This transition accounts for most of the deviation of the total stellar halo mass density profile from a pure power-law distribution of $r^{-3.5}$ at small radii (compare the solid black and dashed green curves in the left panel of Fig.~\ref{fig:rho_profs}).

\begin{figure*}
\includegraphics[width=18cm]{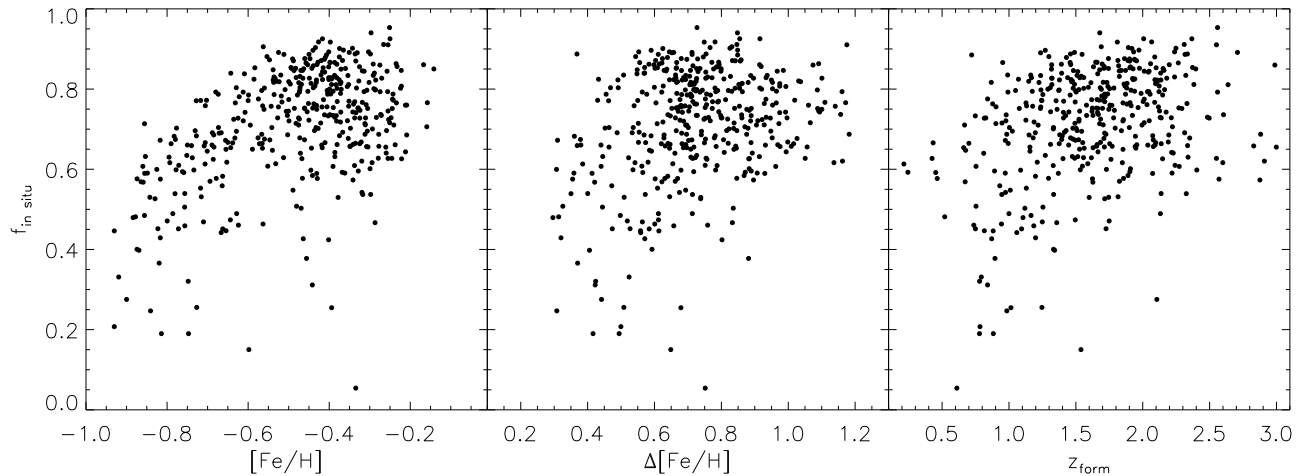}
\caption{\label{fig:frac_insitu} 
The current stellar mass fraction of the bulge+halo component due to in situ star formation as a function of ({\it Left}) mean metallicity, [Fe/H], ({\it Middle}) the metallicity difference between the inner and outer bulge+halo, $\Delta$ [Fe/H] (calculated within and beyond 30 kpc), and ({\it Right})  system formation epoch, $z_{\rm form}$ (i.e., the redshift when half the present-day {\it total} mass (gas+stars+DM) is in place).  Each data point represents a galaxy.  Galaxies with higher in situ mass fractions generally have higher mean metallicities, more prominent metallicity gradients, and are dynamically older, although the system-to-system scatter in these trends is large.
} 
\end{figure*}

Averaged over all of the simulated galaxies, in situ stars account for approximately 68\% of the current mass of the stellar spheroids (with considerable system-to-system scatter, as shown in Fig.~\ref{fig:frac_insitu}), while accreted stars constitute approximately 32\%, with 28\% of the mass formed in satellites prior to accretion, 2\% formed in satellites after accretion, and 2\% accreted smoothly.  Our inferred average in situ fraction is therefore larger than that found recently by \citet{zolotov09}, who found that 3 out of their 4 re-simulated galaxies had in situ mass fractions of $\la20\%$.  However, part of this difference is likely due to the fact that we have made no attempt to distinguish between bulge and halo components (see Section \ref{sec:sample} for a discussion of this point).  Consistent with this is that both \citet{abadi06} and \citet{oser10} find in situ mass fractions of $\sim 40-50\%$ and, like us, these authors have not decomposed the bulge from the halo\footnote{One possible reason why \citet{abadi06} and \citet{oser10} find slightly lower in situ mass fractions than us, is that their simulations ignore metal-line cooling, which is expected to be important.}.  In terms of our simulations, if we limit our comparison to $r > 20$ kpc (well beyond any likely bulge component), the in situ fraction drops to 20\% while the accreted fraction rises to 80\% (with 68\% formed in satellites prior to accretion, 3\% formed in satellites after accretion, and 9\% accreted smoothly).  

In Fig.~\ref{fig:FeH_profile_state} we show the median spherically-averaged stellar metallicity profile split by formation mode.  Here we see that the gradient in the total stellar metallicity profile (solid black curve) is driven by the generally high metallicity of stars formed in situ combined with the fact that the in situ stars dominate the stellar halo by mass in the inner regions.  The metallicity of accreted stars shows no strong variation with radius (except possibly in the central few kpc). The fact that the overall metallicity gradient is established as the result of a transition from in situ-dominated to accretion-dominated stars, appears to resolve the longstanding problem of the inability of purely accretion-based models (e.g., \citealt{font06a,delucia08,cooper10}) to reproduce the strong variation in the mean stellar metallicity with radius observed in M31 (e.g., \citealt{kalirai06,koch08}) and the Milky Way (e.g., \citealt{carollo07,dejong10}).  

In Fig.~\ref{fig:frac_insitu} we show scatter plots of the in situ mass fraction as a function of the mean metallicity of the bulge+halo (left panel) and the metallicity difference between the mean metallicity within and beyond 30 kpc (middle panel).  Each data point represents a simulated galaxy.  Although there is considerable system-to-system scatter, there are identifiable correlations between the in situ mass fraction and the mean metallicity and metallicity difference, in the sense that galaxies with higher in situ mass fractions have higher mean metallicities and more prominent metallicity gradients.  We discuss the right panel of Fig.~\ref{fig:frac_insitu} below.

\subsection{The time of star formation and assembly of the stellar halo}
\label{sec:assembly}

\begin{figure}
\includegraphics[width=\columnwidth]{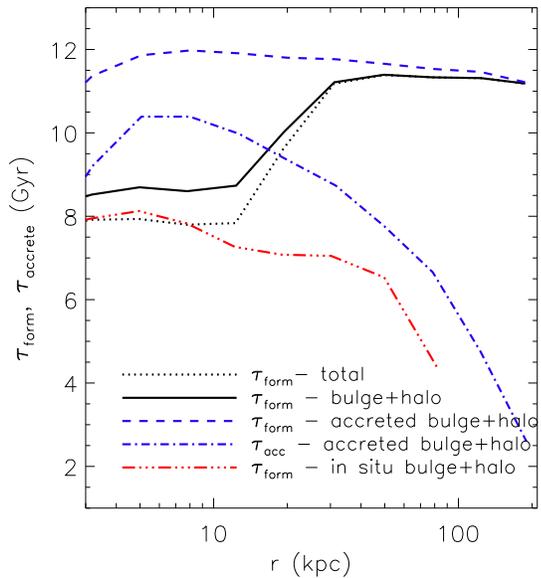}
\caption{\label{fig:tform_profile_state} 
Median spherically-averaged stellar age ($\tau_{\rm form}$) profiles.  The solid black curve represents all stars in the spheroidal component.  The dotted black curve represents all stars (disc+spheroid). The dashed blue curve represents accreted stars in the spheroid.  The triple dot-dashed red curve represents spheroid stars that formed in situ.  The dot-dashed blue curve represents the time of accretion of accreted stars (i.e., the time when the system in which the stars formed became a satellite of the MMP).  The accreted component is old ($\approx 11-12$ Gyr), but the lookback time at which it was assembled varies strongly with radius, with the outer most regions having been accreted quite recently.  The in situ component is younger and was typically formed at $z \approx 1-1.5$.} 
\end{figure}

We now examine the age of the stars and the assembly time of the spheroidal stellar component.  In Fig.~\ref{fig:tform_profile_state} we show profiles of the median spherically-averaged formation lookback time (i.e., the age) of stars, split by formation mode.  Accreted stars (dashed blue curve) are typically quite old ($\approx 11-12$ Gyr), whereas most of the stars that formed in situ and which are located within 30 kpc or so, have a median age of $\approx 7-8$ Gyr.

The finding that stars that were formed in situ are, on average, younger than accreted stars, is consistent with our expectations based on the hierarchical growth of structure.  In particular, stars that formed in situ are, by definition,those that formed in the MMP, whereas accreted stars formed in less massive and typically dynamically older satellites.  For a $1-4 \times 10^{12} M_\odot$ (DM) halo the epoch of formation, defined as the redshift where half the final dark matter mass is in place (\citealt{lacey93}), is $z \approx 1-1.5$ (e.g., \citealt{wechsler02}), which corresponds roughly to the average star formation time of the stars formed in situ.  Of course, baryons feel not only the effects of gravity, but also those of non-gravitational processes such as cooling and heating due to feedback (e.g., from supernovae).  Feedback in particular can result in the delay (or prevention) of significant star formation.  C09 have examined the star formation histories of galaxies in the \gimic\ simulations and found that the peak occurs at $z \sim 1-3$ for normal galaxies, but at $z \sim 3-6$ for dwarf galaxies (which are the systems in which the vast majority of the accreted stars formed).

Returning to Fig.~\ref{fig:frac_insitu}, we plot in the right hand panel the in situ mass fraction of the galaxy against the formation epoch of the system [i.e., the redshift for which the MMP had a total mass due to gas+stars+DM that is half of the final ($z=0$) total mass].  There exists a correlation (again with considerable scatter) between in situ mass fraction and formation epoch, in that older systems have higher in situ mass fractions.  Physically this makes sense, as older systems have been able to form stars quiescently for a longer period of time, while dynamically younger systems accreted relatively larger amounts of mass recently.  The fact that dynamically older systems have higher in situ mass fractions is consistent with the findings of \citet{zolotov09}, although our improved statistics shows that the scatter in this correlation is large.

It is interesting that, even though the accreted stars are quite old (as indicated by the dashed blue curve in Fig.~\ref{fig:tform_profile_state}), many of them were accreted recently, as indicated by the dot-dashed blue curve in Fig.~\ref{fig:tform_profile_state}.  In agreement with hybrid simulations of pure accretion-based halo formation (e.g., \citealt{bullock05,font06a}), we find that the inner accreted halo was formed a long time ago ($\tau_{\rm acc} \approx 9-10$ Gyr), but that the outer accreted halo was assembled as recently as 5 Gyr ago.  We therefore expect to see increased spatial lumpiness and more coherent velocity structure at large radii, since there has been less time since accretion for phase mixing and violent relaxation to act.  We plan to examine the degree of substructure in the spheroidal component in a future paper.

\subsection{Radial migration of in situ stars}
\label{sec:migration}

We have shown that it is in situ star formation that is responsible for the change in the nature of the stellar halo at $r \la 30$ kpc.  Did these stars form at small radii and where then dispersed, as suggested by \citet{zolotov09} (see also \citealt{purcell10})?

\citet{zolotov09} found that the in situ stars in their re-simulations were formed from gas that was brought into the main system in smooth `cold flows'.  In their simulations the stars formed at very small galactocentric distances of $\approx 1$ kpc.  These stars were then dispersed out to larger radii (as far as 20 kpc or so) by $z \approx 2$, with major mergers acting as the primary dispersal mechanism. 

\begin{figure}
\includegraphics[width=\columnwidth]{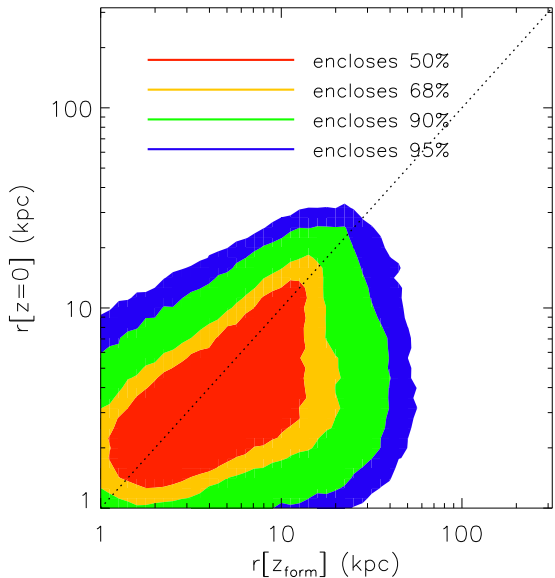}
\caption{\label{fig:rinit_rfinal} Distribution of initial (i.e., at redshift of star formation, $z_{\rm form}$) and final ($z=0$) radii of stars formed in situ.  The red, yellow, green, and blue filled iso-density contours enclose 50\%, 68\%, 90\%, and 95\% of the total number of in situ star particles, respectively.  The dotted curve represents the $r[z=0] = r[z_{\rm form}]$ relation.  On average, in situ stars formed at radii comparable to where they are presently located, although migration in both directions (towards and away from the center) is significant.
} 
\end{figure}

In Fig.~\ref{fig:rinit_rfinal} we show the distribution of initial (i.e., at the time of star formation, $z_{\rm form}$) and final ($z=0$) radii of stars that formed in situ in the \gimic\ simulations.  This distribution is computed by first creating a regular ($50\times50$) 2D grid in initial and final radii (in log space, from $\log_{10}(r) = 0.0-2.5$ in steps of $0.05$) and then counting the number of in situ star particles in each pixel.  The map is normalised by the total number of in situ star particles.  The red, yellow, green, and blue iso-density contours enclose 50\%, 68\%, 90\%, and 95\% of the total number of in situ star particles, respectively.  Fig.~\ref{fig:rinit_rfinal} represents a stacked distribution of all the in situ stars of all simulated galaxies.

Fig.~\ref{fig:rinit_rfinal} shows that the vast majority of in situ stars are presently located at, and were also formed at, relatively small radii ($r \la 30$ kpc).  At these radial distances there is a relatively large scatter about the $r[z=0] = r[z_{\rm form}]$ relation, indicating that in situ stars have migrated in both directions (i.e., toward and away from the present galaxy center).   
A very small fraction of the in situ stars were formed at relatively large radii ($r \ga 30$ kpc).  According to Fig.~\ref{fig:rinit_rfinal} these stars have since migrated inward somewhat.  Naively, it seems surprising that star formation can occur at such large galactocentric radii.  The density and temperature criteria imposed in the simulations for placing gas particles on the star-forming EOS (see Section 2) guarantee that these in situ stars formed from gas that was both dense and cold.  We have analysed the in situ star formation in the outer halo in depth.  First, such star formation is virtually absent in the simulated galaxies at the present day.  Like the majority of the population, these stars formed at $z\ga 1$.  This suggests a physical (rather than spurious numerical) origin to these stars.
Second, stars formed beyond $r[z_{\rm form}] \ga 20$ kpc show up as `streaks' in a plot of $r[z_{\rm form}]$ against $z_{\rm form}$ (i.e., collections of stars formed at very similar time over a wide range of 3D radii), whereas at small radii there is a larger and more homogeneous spread in ages.  This suggests that the large-radii stars are forming in streams (filaments) of dense/cold gas.  Finally, examination of simple movies of the high-resolution simulation shows that the star formation indeed occurs in filaments of cold gas, which has recently been ram pressure-stripped from infalling satellites.  This type of star formation has been seen in SPH simulations previously (e.g., \citealt{puchwein10}).  For idealised test cases it has been shown that the SPH formalism can inhibit the development of fluid instabilities (e.g., \citealt{agertz07,mitchell09}), which could potentially disrupt the clouds/filaments.  Thus, it is possible that our simulations overestimate the degree of in situ star formation at such large radii.  In any case, these stars represent only a small fraction of the in situ population and do not drive trends seen in, e.g., Figs.\ 3-5.

Our results differ from those of \citet{zolotov09} in terms of the radius of in situ star formation, with the in situ stars in our simulations forming at radii comparable to where they are presently located.  Our in situ stellar population is also younger than that of \citet{zolotov09} (who found $z_{\rm form}\approx3$), having formed primarily at $z \sim 1-1.5$, which is similar to that also reported recently by \citet{oser10} for galaxies with masses similar to those in our sample.  The differences in the formation radii and ages of the in situ stars in our simulations and those of \citet{zolotov09} would be expected if the feedback in their simulations is relatively less efficient in low-mass haloes.  This indeed appears to the case, as their simulations significantly overproduce the number of bright satellite galaxies orbiting Milky Way-mass galaxies today (see Fig.\ 4 of \citealt{governato07}).

More importantly, the above differences imply a different formation scenario for  in situ stars in our simulations compared with those of \citet{zolotov09}. In McCarthy et al.\ (in prep.), we will show that a large fraction of the in situ stars formed from the cooling of {\it hot} gas (rather than cold flows, as advocated by \citealt{zolotov09}) during the initial assembly of the disc. This has profound implications for a wide range of properties of these stars today, including their spatial distribution, kinematics and metallicities.

\subsection{A note on ``bulgeless'' galaxies}

\citet{kormendy10} have recently demonstrated, using high quality photometric and spectropscopic observations, that many of the largest nearby disc galaxies lack obvious classical bulge components (i.e., are ``bulgeless'').  These authors have argued that it is difficult to reconcile a significant population of bulgeless galaxies with the standard hierarchical model for galaxy formation, since mergers are a key ingredient of this model and are believed to produce bulges.  However, recent high resolution cosmological simulations have been able to successfully produce bulgeless dwarf galaxies \citep{governato10}.  The key to this success appears to be the implementation of {\it efficient} supernova feedback, which preferentially removes the lowest angular momentum gas via outflows \citep{brook11}.  What about more massive simulated galaxies?  It turns out that the \gimic\ simulations produce a number of massive ``bulgeless'' galaxies, which will be examined in detail in Sales et al.\ (in prep).

For the present study, it is interesting to ask whether the simulated galaxies with small (fractional) bulge components have stellar haloes at all and, if so, what is the contribution of in situ star formation.  It is of interest to note that real bulgeless {\it dwarf} galaxies do indeed have faint, old stellar halos \citep{stinson09} and so do the simulated bulgeless dwarf galaxies of \citet{governato10}.

\begin{figure*}
\includegraphics[width=\columnwidth]{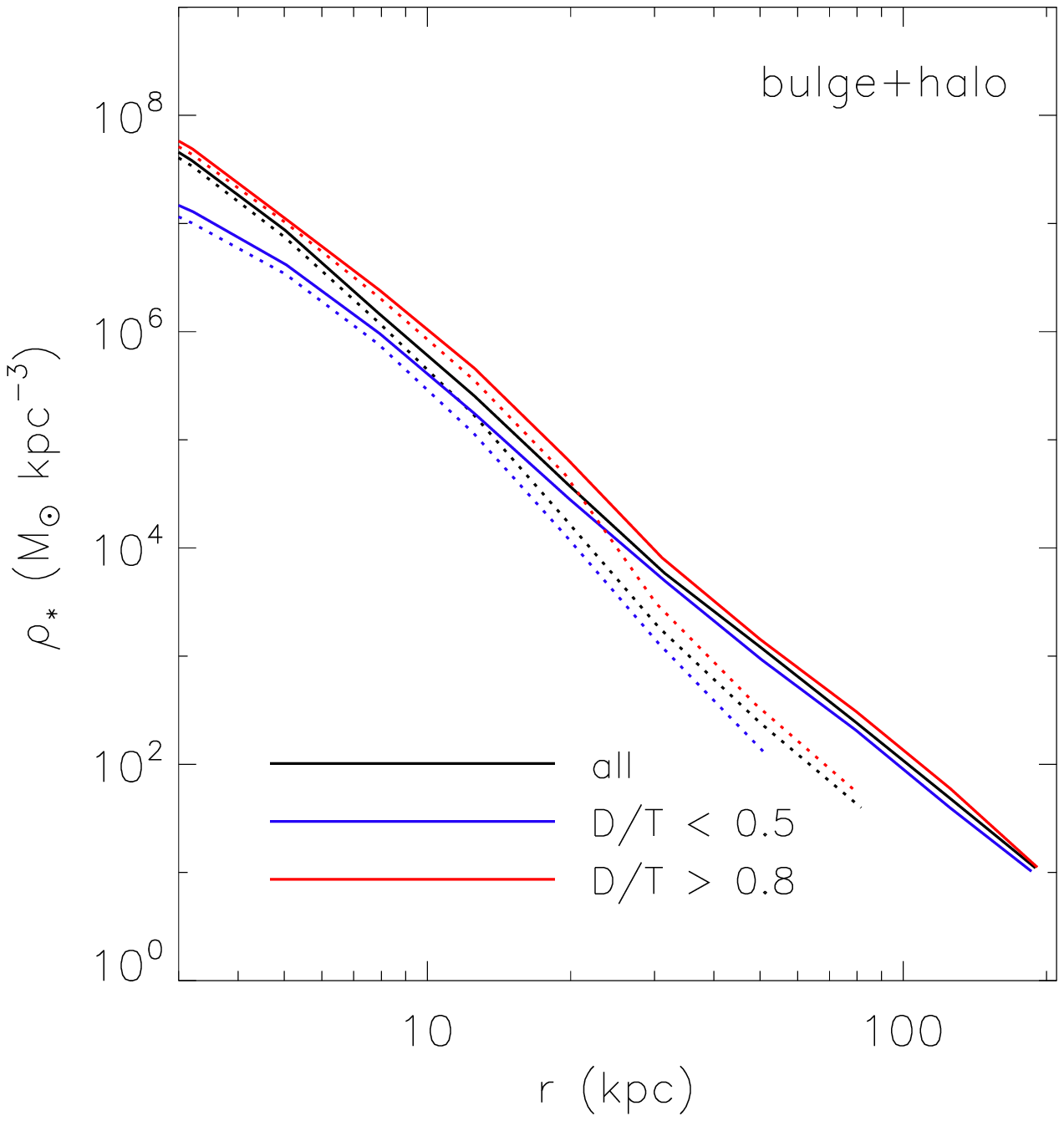}
\includegraphics[width=\columnwidth]{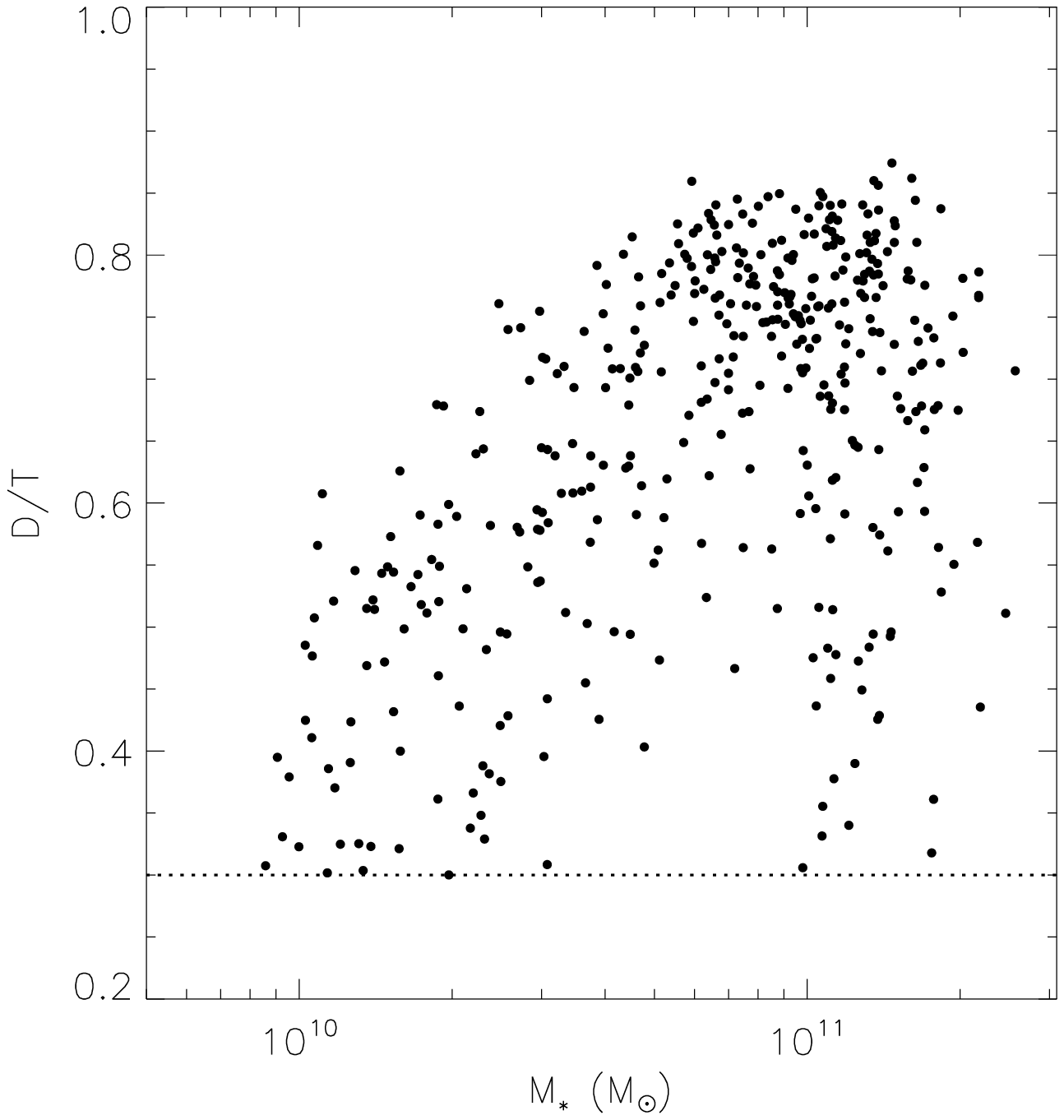}
\caption{\label{fig:bulgeless} {\it Left:} Median spherically-averaged stellar mass density profiles for the total bulge+halo component (solid) curves and just the in situ contribution to the bulge+halo component (dotted) curves.  The black curves represent all 412 simulated galaxies, while the red curves represent only galaxies with `hi' D/T ($> 0.8$) ratios (i.e., small fractional contributions from the bulge+halo) and the blue curves represent only galaxies with 'low' D/T ($< 0.5$) ratios (i.e., have a large fractional contribution from the bulge+halo).  D/T has been computed using the method of C10, described in Section 3.1.  {\it Right:} Scatter plot of D/T versus total stellar mass.  Systems with high D/T ratios actually have more prominent in situ stellar haloes in an absolute sense, since systems with high D/T ratios also tend to have higher total stellar masses.}
\end{figure*}

In the left panel of Fig.~\ref{fig:bulgeless} we compare the median spherically-averaged stellar mass density profiles of the total bulge+halo and the in situ bulge+halo components for our entire sample of galaxies with a subset of galaxies with high D/T and a subset with low D/T.  Interestingly, we find that the total and in situ haloes have {\it larger} mass densities in systems with high D/T ratios.  This is the opposite of what one might have naively expected.  The reason for this trend is provided in the right panel of Fig.~\ref{fig:bulgeless}, which is a scatter plot of D/T versus the total stellar mass (disc+bulge+halo) for all the simulated galaxies.  As can be seen, the highest D/T galaxies have larger total stellar masses (and larger stellar mass fractions) on average than galaxies with low D/T.  This explains why the {\it absolute} mass densities of the in situ bulge+halo component in high-D/T systems is larger than that of this component in low-D/T systems.  It should be noted that this comparison is being made at essentially fixed total system mass ($M_{200}$ varies by only a factor of $4$ over our sample).  By comparison, the total stellar mass $M_*$ varies by factor of $\approx 30$ across the sample.

Thus, based on results in Fig.~\ref{fig:bulgeless} ``bulgeless'' galaxies may actually have the most prominent stellar haloes at fixed total system mass, which is an interesting prediction of the model that should be testable in the future.  The origin of the relatively strong D/T-$M_*$ trend over the narrow range of total system masses considered in this study is unclear and requires further examination.  Furthermore, whether or not this trend is consistent with observations should be investigated.  However, as discussed in Section 3.1, care must be taken to measure D/T in the same way for both real and simulated galaxies.

\section{Summary and Discussion}
\label{sec:concl}

We set out to address the question of the possible dual-nature of the extended stellar distributions around the Milky Way, M31, and other $\sim L_{*}$ nearby galaxies.  In particular, what are the relative contributions (by mass) to the stellar spheroid from tidal stripping/disruption of satellites and `in situ' star formation?  And what is the origin of the metallicity gradients observed in the Milky Way and M31?

We have used the \gimic\ suite of cosmological hydrodynamical simulations to address these important questions.  \gimic\ is ideal for this purpose as it contains large numbers ($\approx 400$) of Milky Way-mass galaxies simulated at relatively high-resolution ($M_{g} \simeq 1.16 \times 10^{7} h^{-1}$ M$_{\odot}$ and a gravitational softening of $1 h^{-1}$ kpc) and with sophisticated prescriptions for radiative cooling, chemodynamics, and supernova feedback.

Our main conclusions are as follows:
\begin{itemize}
\item{The simulated stellar mass density of the spheroid (bulge+halo) falls off as $r^{-3.5}$ at large radii ($r \ga 30$ kpc), in agreement with the findings of previous theoretical studies based on hybrid analytic+N-body methods and cosmological hydrodynamical re-simulations, as well as deep observations of M31.  Fitting all $\approx 400$ simulated galaxies separately, we find the 1-sigma (2-sigma) scatter in the logarithmic slope is $\approx 0.4$ ($\approx 0.6$).}
\item{In the inner region, the simulated mass density profile can be fit very well with a de Vaucouleurs profile with a median $r_{\rm eff} \approx 1.8$ kpc (with considerable system-to-system scatter), and relatively well with a power-law with logarithmic slope $\approx -3.0 \pm -0.4$.  This is shallower than in the outer regions, but note that the mass density profile has a higher normalisation at small radii, so that a power-law with slope $-3.5$ extrapolated to the inner regions underpredicts the mass density there.  The increased normalisation and still fairly steep profile at small radii differs from that found previously with hybrid analytic+N-body methods, which is most likely due to their neglect of in situ star formation.  The shape of our simulated mass density profile matches that inferred for M31 (see Fig.~\ref{fig:sig_profs}) and other nearby galaxies remarkably well.}
\item{{\it Significant negative metallicity gradients are a ubiquitous feature of the simulated galaxy populations.}  Typically [Fe/H] drops by 0.6 to 0.9 dex from the inner ($r < 30$ kpc) to the outer ($r \approx 30-100$ kpc) regions (see middle panel of Fig.~\ref{fig:frac_insitu}).  This drop is compatible with that inferred for the MW (e.g., \citealt{dejong10}) and M31 (see Fig.~\ref{fig:iron_profs}).}

\item{On average, 2/3 of the mass of the stellar spheroid (bulge+halo) is composed of stars that formed in situ, but with considerable system-to-system scatter.  Limiting the comparison to $r > 20$ kpc, this fraction drops to 1/5 on average.}
\item{The large contribution from in situ star formation in the inner regions drives the change in slope of the mass density profile at $r \simeq 30$~kpc and also gives rise to large-scale metallicity gradients.}
\item{On average, the in situ stars were formed at radii comparable to their present distance from the Galactic center, with considerable migration in both directions (towards and away from the center; see Fig.~\ref{fig:rinit_rfinal}).  Typically, the stars were formed at $z \approx 1-1.5$, in a proto-disc.  See McCarthy et al.\ (in prep.) for further discussion.}
\item{At fixed total mass, $M_{200}$, galaxies with high D/T ratios actually have the most prominent stellar haloes (in terms of absolute mass density).  This owes to the presence of a strong trend between D/T and total (disc+bulge+halo) stellar mass.}
\end{itemize}

In the Introduction we noted that the large differences in the properties of the stellar bulge+halo of the Milky Way and M31 (particularly their metallicities) has raised questions about the compatibility with the $\Lambda$CDM scenario.  As we have shown, however, there is considerable system-to-system scatter in the mean metallicities (as well as in the prominence of metallicity gradients) of the simulated Milky Way-mass systems (see Fig.~\ref{fig:frac_insitu}). The scatter is roughly compatible with the observed differences between the Milky Way and M31.  

M31 is typically considered to be more dynamically active than the Milky Way (\citealt{hammer07,hammer10}). The correlation between the epoch of formation and the in situ mass fraction found in the simulations would therefore suggest that the Milky Way has a higher in situ mass fraction than M31 (as also argued by \citealt{zolotov09}).  However, this is difficult to reconcile with the fact that we find that the in situ mass fraction is also correlated with system metallicity, in the sense that more metal rich systems have higher in situ fractions (and that M31 has a more metal rich spheroidal component than the Milky Way).  However, as we have already noted, the simulations predict a large degree of scatter in these correlations, which may reconcile this issue.  

It would clearly be very interesting to {\it measure} the in situ mass fraction of the spheroidal components of the Milky Way and Andromeda and compare it directly to the predictions of the simulations (e.g., in Fig.~\ref{fig:frac_insitu}).  Given their different formation histories, we expect there to be differences in the detailed spatial distributions and kinematics of the in situ and accreted stellar components, and this may hold the key to unambiguously separating these components from one another.  We examine the kinematic properties, as well as the 3D spatial structure, of the in situ and accreted components in McCarthy et al.\ (in prep.).

It should be borne in mind that there are clear limitations in comparing a large sample of simulated galaxies (which could have unforseen ``systematic errors'', e.g., lack important physics) to one with very poor number statistics.  Probing the diffuse stellar haloes of large samples of more distant galaxies will go a very long way towards verifying (or refuting) the picture we have proposed.  At present, it is only possible to probe the haloes of distant galaxies by stacking large samples of them (e.g, \citealt{zibetti04}), but then one loses information about the system-to-system scatter, which is predicted to be large.  In the coming years, observations with the Large Synoptic Survey Telescope (for example), however, will allow us to probe the outer regions of distant galaxies in exquisite detail and will provide an excellent testbed for simulations of galaxy formation.

\section*{Acknowledgements}
 
We thank the referee, Fabio Governato, for suggestions that improved the paper.
We thank Raja Guhathakurta, Karrie Gilbert, Jason Kalirai, Michael Rich, Andreas Koch, Alan McConnachie, Scott Chapman and Qi Guo for useful discussions. ASF is supported by a Royal Society Dorothy Hodgkin Fellowship at the University of Cambridge. IGM is supported by a Kavli Institute Fellowship at the University of Cambridge.

\begin{figure*}
\includegraphics[width=\columnwidth]{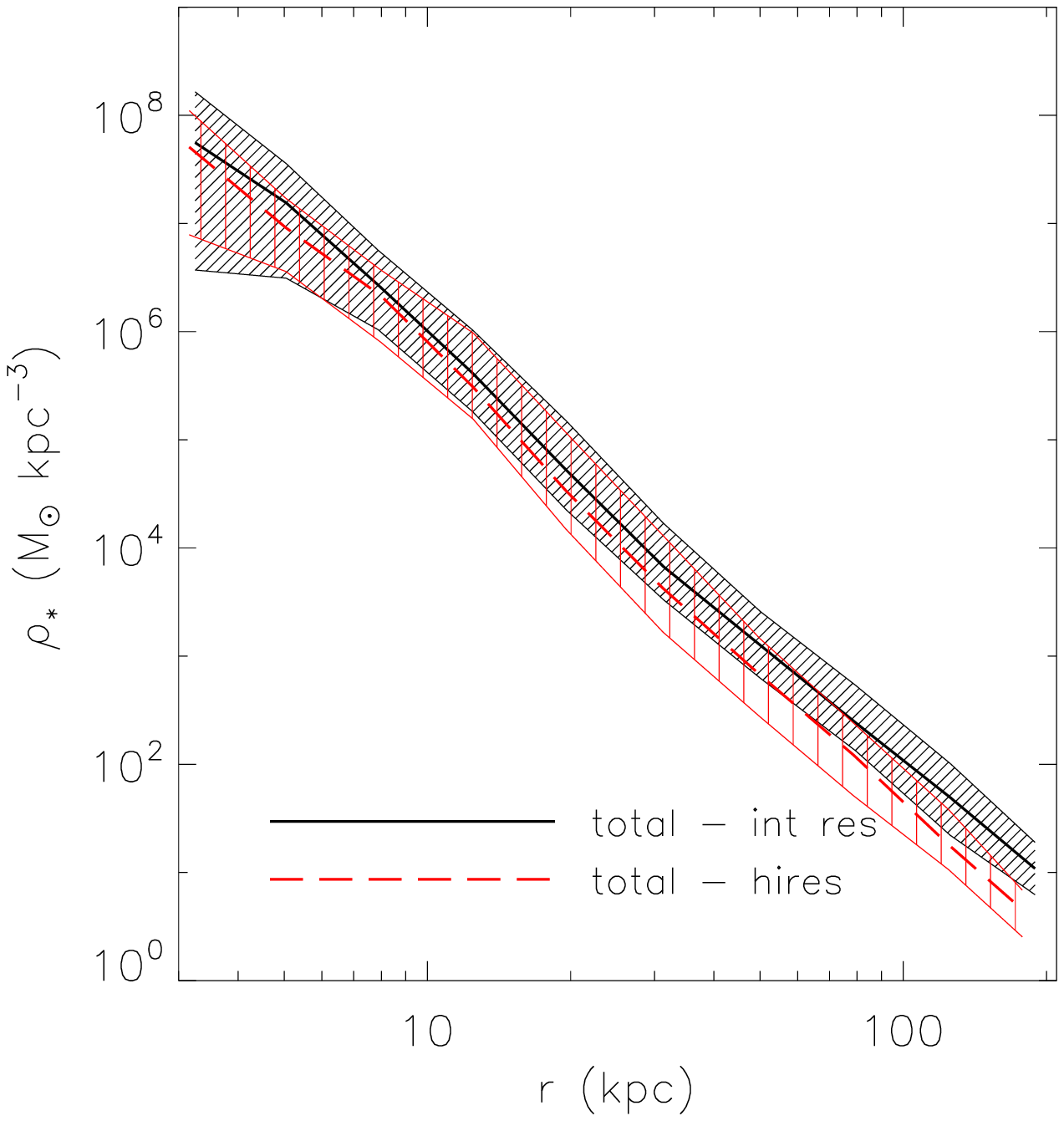} 
\includegraphics[width=\columnwidth]{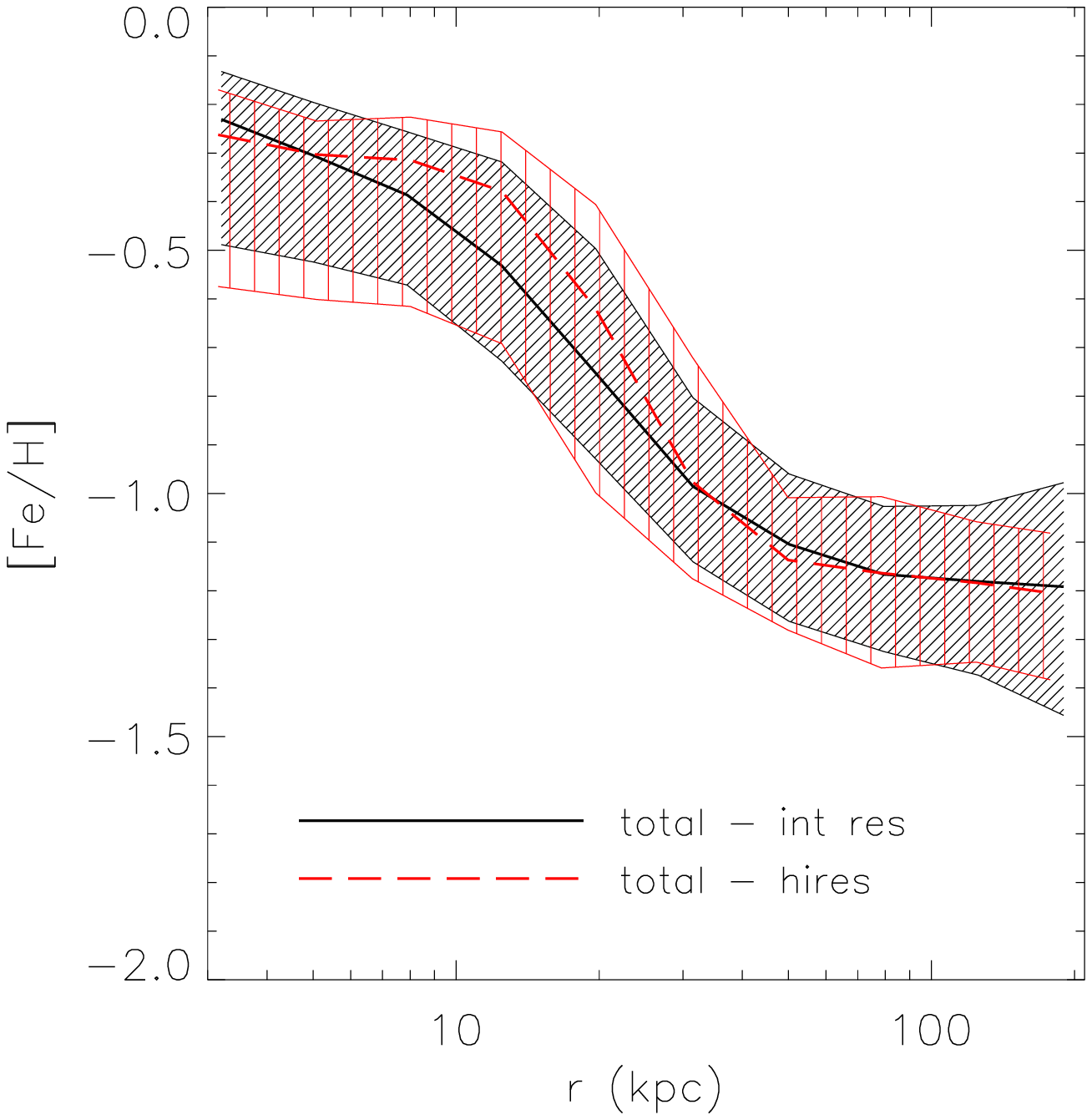}
\caption{\label{fig:rho_profs_res} {\it Left:} Median spherically-averaged stellar mass density profiles for all Milky Way-mass systems in the high- and intermediate-resolution simulations. The scatter around the mean is indicated in each case by the different hashed regions (thinly spaced oblique lines for the intermediate-resolution run and widely spaced vertical lines for the high-resolution run).  {\it Right:} Median spherically-averaged stellar metallicity profiles for the two runs. Overall, the agreement between the high-resolution and intermediate-resolution \gimic\ simulations is good for both the mass density and metallicity profiles.
}\end{figure*}

\begin{figure}
\includegraphics[width=\columnwidth]{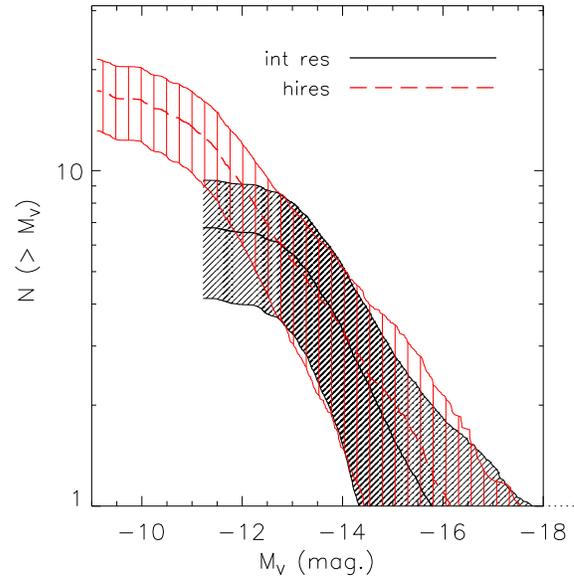} 
\caption{\label{fig:sat_LF_res} The mean satellite luminosity function of simulated galaxies in the intermediate-resolution (black) and high-resolution (red) \gimic\ simulation.  The hatched regions represent the Poisson error distributions for these luminosity function.  For $M_V \la -12.5$ there is good agreement.
}\end{figure}

\section*{Appendix: Numerical convergence}

Here we investigate the sensitivity of our results to numerical resolution.  As noted in Section 2, the $-2\sigma$ sphere has been simulated at eight times better mass resolution than the intermediate-resolution runs analysed in the present study.  

In Fig.~\ref{fig:rho_profs_res} we compare the mass density and metallicity profiles (median + scatter) of 50 Milky Way-mass disc galaxies in the high-resolution simulation with those of the $\approx 400$ galaxies in the intermediate-resolution runs.  
Overall, the agreement between the high-resolution and intermediate-resolution \gimic\ simulations is good for both the mass density and metallicity profiles.

We have also quantified the mass and fractional contribution of the in situ population to the spheroid in the high-resolution and intermediate-resolution simulations.  The mean (median) mass of the in situ component is $2.68$ ($2.48$) $\times 10^{10} M_\odot$ and $2.01$ ($2.06$) $\times 10^{10} M_\odot$ in the intermediate- and high-res runs, respectively.  Star formation is therefore slightly less efficient in the high-res run (as can also been seen by the small difference in the normalisation of the density profiles in the top left panel of Fig.~\ref{fig:rho_profs_res}).  The {\it fractional} contribution of the in situ population to the bulge+halo has a mean (median) of $0.71$ ($0.73$) for the intermediate-resolution runs and has a mean (median) of $0.69$ ($0.74$) for the high-res run.

In Fig.~\ref{fig:sat_LF_res} we compare the mean satellite LF for 50 Milky Way-mass disc galaxies in the high-resolution simulation with those of the $\approx 400$ galaxies in the intermediate-resolution runs.  For $M_V \la -12.5$ there is good convergence.


\begin{thebibliography}{99}

\bibitem[Abadi et al.(2003)]{abadi03} Abadi, M.~G., Navarro, 
J.~F., Steinmetz, M., \& Eke, V.~R.\ 2003, ApJ, 591, 499 

\bibitem[Abadi et al.(2006)]{abadi06} Abadi, M.~G., Navarro, J.~F., Steinmetz,~M. 2006, MNRAS, 365, 747

\bibitem[Agertz et al.(2007)]{agertz07} Agertz, O., et al.\ 
2007, MNRAS, 380, 963 

\bibitem[Asplund et al.(2005)]{asplund05} Asplund, M., Grevesse, N. Sauval, A. J. 2005, ASPC, 336, 25

\bibitem[Battaglia et al.(2005)]{battaglia05} Battaglia, G. et al. 2005, MNRAS, 364, 433

\bibitem[Beers et al.(2011)]{beers11} Beers, T.~C., et al.\ 
2011, arXiv:1104.2513 

\bibitem[Bell et al.(2008)]{bell08} Bell, E.~F., et al.\ 2008, ApJ, 680, 295 

\bibitem[Bell  \& de Jong(2001)]{bell01} Bell, E.~F., de Jong, R.~S.\ 2001, ApJ, 550, 212

\bibitem[Blumenthal et al.(1986)]{blumenthal86} Blumenthal, G. R., Faber, S. M., Flores, R., Primack, J. R. 1986, ApJ, 301, 27

\bibitem[Booth \& Schaye(2009)]{booth09} Booth, C.~M., \& Schaye, J.\ 2009, MNRAS, 398, 53 
 
\bibitem[Boylan-Kolchin et al.(2010)]{boylan-kolchin10} Boylan-Kolchin, M., Springel, V., White, S. D. M., Jenkins, A. 2010, MNRAS, 406, 896

\bibitem[Bower et al.(2006)]{bower06} Bower, R. G. et al., 2006, MNRAS, 370, 645 
\bibitem[Brook et al.(2004)]{brook04} Brook, C. B., Kawata, D., Gibson, B. K., Flynn, C. 2004, MNRAS, 349, 52

\bibitem[Brook et al.(2011)]{brook11} Brook, C.~B., et al.\ 
2011, MNRAS, 595 

\bibitem[Brooks et al.(2007)]{brooks07} Brooks, A.~M., 
Governato, F., Booth, C.~M., Willman, B., Gardner, J.~P., Wadsley, J., 
Stinson, G., \& Quinn, T.\ 2007, ApJ, 655, L17 

\bibitem[Brown et al(2006a)]{brown06a} Brown, T.~M.. et al 2006a, ApJL, 636, 89

\bibitem[Brown et al(2006b)]{brown06b} Brown, T.~M.. et al 2006b, ApJ, 652, 323

\bibitem[Brown et al(2007)]{brown07} Brown, T.~M.. et al 2007, ApJ, 658, L95

\bibitem[Bruzual \& Charlot(2003)]{bruzual03} Bruzual, G.  Charlot, S. 2003, MNRAS, 344, 1000

\bibitem[Bullock \& Johnston(2005)]{bullock05} Bullock, J.~S., Johnston, K.~V. 2005, ApJ, 635, 931 

\bibitem[Carollo et al.(2007)]{carollo07} Carollo, D. et al. 2007, Nature, 7172, 1020

\bibitem[Carollo et al.(2010)]{carollo10} Carollo, D., Beers, T. C., Chiba, M. et al. 2010, ApJ, 712, 692

\bibitem[Chabrier(2001)]{chabrier01} Chabrier, G. 2001, ApJ, 554, 1274

\bibitem[Chabrier(2003)]{chabrier03} Chabrier, G. 2003, PASP, 115, 763

\bibitem[Chen et al.(2001)]{chen01} Chen, B. et al. 2001, ApJ, 553, 184
 
\bibitem[Chiba \& Beers(2000)]{chiba00} Chiba, M., Beers, T. 2000, AJ,  119, 2843

\bibitem[Chapman et al.(2006)]{chapman06} Chapman, S.~C., Ibata, 
R., Lewis, G.~F., Ferguson, A.~M.~N., Irwin, M., McConnachie, A., 
\& Tanvir, N.\ 2006, ApJ, 653, 255 

\bibitem[Christensen et al.(2010)]{christensen10} Christensen, C.~R., 
Quinn, T., Stinson, G., Bellovary, J., \& Wadsley, J.\ 2010, ApJ, 717, 121 

\bibitem[Cooper et al.(2010)]{cooper10} Cooper, A. P. et al. 2010, MNRAS, 406, 744

\bibitem[Crain et al.(2009)]{crain09} Crain, R. A. et al. 2009, MNRAS, 399, 1773 (C09)

\bibitem[Crain et al.(2010)]{crain10} Crain, R. A., McCarthy, I. G., Frenk, C. S., Theuns, T., Schaye, J. 2010, MNRAS, 407, 1403 (C10)

\bibitem[Croton et al.(2006)]{croton06} Croton, D. J. et al. 2006, MNRAS, 367, 864

\bibitem[Dalla Vecchia \& Schaye(2008)]{dallavecchia08} Dalla Vecchia, C. Schaye, J. 2008, MNRAS, 387, 1431

\bibitem[Deason et al.(2011a)]{deason11a} Deason, A.~J., 
Belokurov, V., \& Evans, N.~W.\ 2011, MNRAS, 411, 1480 

\bibitem[Deason et al.(2011b)]{deason11b} Deason, A.~J., 
Belokurov, V., \& Evans, N.~W.\ 2011, MNRAS, submitted (arXiv:1104.3220)

\bibitem[Deason et al.(2011c)]{deason11c} Deason, A.~J., et al.\ 2011, MNRAS, in press (arXiv:1101.0816)

\bibitem[Dehnen 
\& Binney(1998)]{dehnen98} Dehnen, W., \& Binney, J.\ 1998, MNRAS, 294, 429 

\bibitem[de Jong et al.(2007)]{dejong07} de Jong, R. S., et al. 2007 IAUS, 241, 503

\bibitem[de Jong et al.(2010)]{dejong10} de Jong, J. T. A, Yanny, B., Rix, H.-W., Dolphin, A. E., Martin, N., Beers, T. C. 2010, ApJ, 714, 663

\bibitem[De Lucia \& Helmi(2008)]{delucia08} De Lucia, G., Helmi, A. 2008, MNRAS, 391, 14 

\bibitem[Diemand et al.(2005)]{diemand05} Diemand, J., Madau, P., Moore, B. 2005, MNRAS, 364, 367

\bibitem[Diemand, Kuhlen \& Madau (2007)]{diemand07} Diemand, J., Kuhlen, M., Madau, P. 2007, ApJ, 657, 262

\bibitem[Dolag et al.(2009)]{dolag09} Dolag, K., Borgani, S., Murante, G., Springel, V. 2009, MNRAS, 399, 497

\bibitem[Duffy et al.(2010)]{duffy10} Duffy, A. R., Schaye, J., Kay, S. T., Dalla Vecchia, C., Battye, R. A., Booth, C. M., 2010, MNRAS, 405, 2161

\bibitem[Eggen, Lynden-Bell \& Sandage(1962)]{eggen62} Eggen, O., J.
Lynden-Bell, D.  Sandage, A. R. 1962, ApJ, 136, 748

\bibitem[Evans \& Wilkinson(2000)]{evans00} Evans, N. W., Wilkinson, M. I. 2000, MNRAS, 316, 929
 
\bibitem[Ferguson et al.(2002)]{ferguson02} Ferguson, A.~M.~N., 
Irwin, M.~J., Ibata, R.~A., Lewis, G.~F.,  Tanvir, N.~R.\ 2002, AJ, 124, 
1452 

\bibitem[Ferland et al.(1998)]{ferland98} Ferland, G. J., Korista,
K. T., Verner, D. A., Ferguson, J. W., Kingdon, J. B., Verner,
E. M. 1998, PASP, 110, 761

\bibitem[Flynn et al.(2006)]{flynn06} Flynn, c., Holmberg, J., Portinari, L., Fuchs, B., Jahrei\ss, H. 2006, MNRAS, 372, 1149

\bibitem[Font et al.(2006a)]{font06a} Font, A.~S., Johnston, 
K.~V., Bullock, J.~S.,  Robertson, B.~E.\ 2006a ApJ, 638, 585

\bibitem[Font et al.(2006b)]{font06b} Font, A.~S., Johnston, 
K.~V., Bullock, J.~S.,  Robertson, B.~E.\ 2006b, ApJ, 646, 886 

\bibitem[Font et al.(2006c)]{font06c} Font, A.~S., Johnston, K.~V., Guhathakurta, P., Majewski, S. R., Rich, R. M. \ 2006c, AJ, 131, 1436

\bibitem[Font et al.(2008)]{font08} Font, A.~S., Johnston, 
K.~V., Ferguson, A.~M.~N., Bullock, J.~S., Robertson, B.~E., Tumlinson, J., 
 Guhathakurta, P.\ 2008, ApJ, 673, 215 

\bibitem[Gao et al.(2008)]{gao08} Gao, L. et al. 2008, MNRAS, 387, 536

\bibitem[Geehan et al.(2006)]{geehan06} Geehan, J.~J., Fardal, 
M.~A., Babul, A., \& Guhathakurta, P.\ 2006, MNRAS, 366, 996 

\bibitem[Gibson et al.(2007)]{gibson07} Gibson, B. K., S{\'a}nchez-Bl{\'a}zquez, P., Courty, S., Kawata, D. 2007, EAS Publication Series, 24, 133

\bibitem[Gil de Paz et al.(2007)]{gildepaz07} Gil de Paz, A., et 
al.\ 2007, ApJS, 173, 185 

\bibitem[Gilbert et al.(2006)]{gilbert06} Gilbert, K. M. et al. 2006, ApJ, 652, 1188

\bibitem[Gilbert et al.(2007)]{gilbert07} Gilbert, K. M. et al. 2007, ApJ, 668, 245

\bibitem[Gilbert et al.(2009a)]{gilbert09a} Gilbert, K. M., Font, A. S., Johnston, K. V., Guhathakurta, P. 2009a, ApJ, 701, 776

\bibitem[Gilbert et al.(2009b)]{gilbert09b} Gilbert, K. M. et al. 2009b, ApJ, 705, 1275

\bibitem[Gnedin et al.(2004)]{gnedin04} Gnedin, O. Y., Kravtsov, A. V., Klypin, A. A., Nagai, D. 2004, ApJ, 616, 16

\bibitem[Governato et al.(2007)]{governato07} Governato, F., Willman, B., Mayer, L ., Brooks, A., Stinson, G., Valenzuela, O., Wadsley, J., Quinn, T. 2007, MNRAS, 374, 1479
 
\bibitem[Governato et al.(2010)]{governato10} Governato, F., et 
al.\ 2010, Nature, 463, 203 

\bibitem[Guhathakurta et al.(2005)]{guhathakurta05} Guhathakurta, P., Ostheimer, J.~C., Gilbert, K.~M., Rich,~R.~M., Majewski, S.~R., Kalirai, J. S., Reitzel, D. B.,  Patterson, R. J.  2005, arXiv:astro-ph/0502366

\bibitem[Guo et al.(2010)]{guo10} Guo, Q., White, S. D. M., Li, C., Boylan-Kolchin, M., 2010, MNRAS, 404, 1111

\bibitem[Haardt \& Madau(2001)]{haardt01} Haardt, F. \& Madau, P. 2001, in Clusters of galaxies and the high redshift universe observed in X-rays, ed. D.M. Neumann \& J.T.T. Van.

\bibitem[Hammer et al.(2007)]{hammer07} Hammer, F., Puech, M., Chemin, L., Flores, H., Lehnert, M. D. 2007, ApJ, 662, 322

\bibitem[Hammer et al.(2010)]{hammer10} Hammer, F., Yang, Y. B., Wang, J. L., Puech, M., Fores, H., Fouquet, S. 2010,  ApJ, 725, 542

\bibitem[Harris(1976)]{harris76} Harris, W. E. 1976, AJ, 81, 1095

\bibitem[Hopkins et al.(2009)]{hopkins09} Hopkins, P.~F., Cox, 
T.~J., Younger, J.~D., \& Hernquist, L.\ 2009, ApJ, 691, 1168 

\bibitem[Ibata et~al.(2007)]{ibata07} Ibata, R., Martin, N.~F., Irwin, M., Chapman, S., Ferguson, A.~M.~N., Lewis, G.~F.,  McConnachie, A.~W. \ 2007,
ApJ, 671, 1591

\bibitem[Ibata et~al.(2009)]{ibata09} Ibata, R., Mouhcine, M.  Rejkuba, M. 2009, MNRAS, 395, 126

\bibitem[Irwin et al.(2005)]{irwin05} Irwin, M.~J., Ferguson, A.~M.~N., Ibata, R.~A., Lewis,~G.~F.,  Tanvir,~N.~R. 2005, ApJ, 628, L105

\bibitem[Ivezi\'c et al.(2000)]{ivezic00} Ivezi\'c, \v{Z}. et al. 2000, AJ, 120, 963

\bibitem[Ivezi\'c et al.(2008)]{ivezic08} Ivezi\'c, \v{Z}. et al. 2008, ApJ, 684, 287

\bibitem[Johnston et al.(2008)]{johnston08} Johnston, K. V., Bullock, J. S., Sharma, S., Font, A. S., Robertson, B.  Leitner, S. N., 2008, ApJ, 689, 936

\bibitem[Jur{\'i}c et al.(2008)]{juric08} Jur{\'i}c, M. et al. 2008, ApJ, 673, 864

\bibitem[Kalirai et al.(2006)]{kalirai06} Kalirai, J.~S. et al. \ 2006, ApJ, 648, 389

\bibitem[Kalirai et al.(2009)]{kalirai09} Kalirai, J. S. et al. 2009, arXiv:0911.1998v1

\bibitem[Karanchentsev \& Kashibadze(2006)]{karanchentsev06} Karachentsev, I. D.; Kashibadze, O. G. (2006). Astrophysics, 49, 3

\bibitem[Kennicutt(1998)]{kennicutt98} Kennicutt, R. C., Jr. 1998, ApJ, 498, 541

\bibitem[Kobayashi(2004)]{kobayashi04} Kobayashi, C. 2004, MNRAS, 347, 740

\bibitem[Kobayashi(2005)]{kobayashi05} Kobayashi, C. 2005, MNRAS, 361, 1216

\bibitem[Koch et al.(2008)]{koch08} Koch, A. et al. 2008, ApJ, 689,958

\bibitem[Kochanek(1996)]{kochanek96} Kochanek, C. S. 1996, ApJ, 457, 228

\bibitem[Komatsu et al.(2009)]{komatsu09} Komatsu, E. et al. 2009, ApJS, 180, 330

\bibitem[Kormendy et al.(2010)]{kormendy10} Kormendy, J., Drory, 
N., Bender, R., \& Cornell, M.~E.\ 2010, ApJ, 723, 54 

\bibitem[Kroupa(1998)]{kroupa98} Kroupa, P. 1998, in Rebolo, R., Martin, E. L., Osorio, M. R. Z., eds, ASP COnf. Ser. 134, Brown Dwarfs and Extrasolar Planets, Astron. Soc. Pac., San Francisco, p483

\bibitem[Lacey \& Cole(1993)]{lacey93} Lacey, C., Cole, S, 1993, MNRAS, 262, 627

\bibitem[Mart{\'i}nez-Delgado et al.(2010)]{martinez-delgado10} Mart{\'i}nez-Delado, D. et al. 2010, AJ, 140, 962 

\bibitem[Mateo(1998)]{mateo98} Mateo, M. L. 1998, ARAA, 36, 435

\bibitem[McCarthy et al.(2008)]{mccarthy08} McCarthy, I. G., Frenk, C. S., Font, A. S., Lacey, C. G., Bower, R. G., Mitchell, N. L., Balogh, M. L., Theuns, T. 2008, MNRAS, 383, 593

\bibitem[McConnachie \& Irwin(2006)]{mcconnachie06} McConnachie, A. W.  Irwin, M. J. 2006, MNRAS, 365, 1263

\bibitem[McConnachie et al.(2009)]{mcconnachie09} McConnachie, A. W. et al. 2009, Nature, 461, 66

\bibitem[Mitchell et al.(2009)]{mitchell09} Mitchell, N.~L., 
McCarthy, I.~G., Bower, R.~G., Theuns, T., 
\& Crain, R.~A.\ 2009, MNRAS, 395, 180 

\bibitem[Morrison et al.(2000)]{morrison00} Morrison, H., Mateo, M., Olszewski, E., Harding, P., Donm-Palmer, R., Freeman, K., Norris, J., Morita, M.  2000, AJ, 119, 2254

\bibitem[Mouhcine et al.(2005a)]{mouhcine05a} Mouhcine, M., Ferguson, H. C., Rich, R. M., Brown,T. M.  Smith, T. E. 2005a, ApJ, 633, 821

\bibitem[Mouhcine et al.(2010)]{mouhcine10} Mouhcine, M., Ibata, 
R., \& Rejkuba, M.\ 2010, ApJL, 714, L12 

\bibitem[Mouhcine et al.(2005b)]{mouhcine05b} Mouhcine, M., Rich, R. M.,  Ferguson, H. C., Brown,T. M.  Smith, T. E. 2005b, ApJ, 633, 828

\bibitem[Navarro, Frenk \& White(1996)]{navarro96} Navarro, J.~F., Frenk, 
C.~S.,  White, S.~D.~M.\ 1996, ApJ, 462, 563 

\bibitem[Nissen \& Schuster(2010)]{nissen10} Nissen, P. E., Schuster, W. J. 2010, A\&A, 511, L10

\bibitem[Okamoto et al.(2005)]{okamoto05} Okamoto, T., Eke, V. R., Frenk, C. S., Jenkins, A. 2005, MNRAS, 363, 1299

\bibitem[Oser et al.(2010)]{oser10} Oser, L., Ostriker, J.~P., 
Naab, T., Johansson, P.~H., \& Burkert, A.\ 2010, ApJ, submitted (arXiv:1010.1381)

\bibitem[Preston, Shectman  Beers(1991)]{preston91} Preston, G. W., Shectman, S. A.  Beers, T. C. 1991, ApJ, 375, 121

\bibitem[Pritchet \& van den Bergh(1994)]{pritchet94} Pritchet, C. J., van den Bergh, S. 1994, AJ, 107, 1730

\bibitem[Puchwein et al.(2010)]{puchwein10} Puchwein, E., 
Springel, V., Sijacki, D., \& Dolag, K.\ 2010, MNRAS, 406, 936 

\bibitem[Purcell et al.(2010)]{purcell10} Purcell, C. W., Bullock, J. S., Kazantzidis, S. 2010, MNRAS, 404, 1711

\bibitem[Reid et al.(2009)]{reid09} Reid, M. J. et al. 2009, ApJ, 700, 137.

\bibitem[Richardson et al.(2009)]{richardson09} Richardson, J.~C., 
et al.\ 2009, MNRAS, 396, 1842 

\bibitem[Saha(1985)]{saha85} Saha, A. 1985, ApJ, 289, 310

\bibitem[Sales et al.(2010)]{sales10} Sales, L.~V., Navarro, 
J.~F., Schaye, J., Vecchia, C.~D., Springel, V., 
\& Booth, C.~M.\ 2010, MNRAS, 409, 1541 

\bibitem[Scannapieco et al.(2010)]{scannapieco10} Scannapieco, C., 
Gadotti, D.~A., Jonsson, P., \& White, S.~D.~M.\ 2010, MNRAS, 407, L41 

\bibitem[Schaye \& Dalla Vecchia(2008)]{schaye08} Schaye, J., Dalla Vecchia, C. 2008, MNRAS, 383, 1210

\bibitem[Schaye (2004)]{schaye04} Schaye, J. 2004, ApJ, 609, 667

\bibitem[Schaye et al.(2010)]{schaye10} Schaye, J., et al. 2010, MNRAS, 402, 1536

\bibitem[Schoenrich et al.(2010)]{schoenrich10} Schoenrich, R., 
  Asplund, M., \& Casagrande, L.\ 2010, MNRAS, submitted (arXiv:1012.0842)

\bibitem[Searle \& Zinn(1978)]{searle78} Searle, L., Zinn, R. 1978, ApJ, 225, 357 

\bibitem[Sesar et al.(2011)]{sesar11} Sesar, B., Juric, M., \& Ivezic, Z.\ 2011, ApJ, 731, 4 

\bibitem[Siegel et al.(2002)]{siegel02} Siegel, M., Majewski, S., Reid, I., Thompson, I. 2002, ApJ, 578, 151

\bibitem[Smith et al.(2009)]{smith09} Smith, M. C., et al., 2009, MNRAS, 399, 1223

\bibitem[Springel et al.(2001)]{springel01} Springel, V., White, S. D. M., Tormen, G., Kauffman, G. 2001, MNRAS, 328, 726

\bibitem[Springel (2005)]{springel05} Springel, V. 2005, MNRAS, 364, 1105

\bibitem[Springel et al.(2005a)]{springel_etal_05a} Springel, V. et al. 2005a, Nature, 435, 629

\bibitem[Springel et al.(2005b)]{springel_etal_05b} Springel, V., Di 
Matteo, T., \& Hernquist, L.\ 2005b, MNRAS, 361, 776 

\bibitem[Springel et al.(2008)]{springel08} Springel, V. et al. 2008, MNRAS, 391, 1685

\bibitem[Starkenburg et al.(2009)]{starkenburg09} Starkenburg, E. et al. 2009, ApJ, 698, 567

\bibitem[Stewart et al.(2008)]{stewart08} Stewart,~K.~R., Bullock, J.~S., Wechsler,~R.~H.,  Maller, A.~H., Zentner, A.~R., 2008, ApJ, 683, 597

\bibitem[Stinson et al.(2009)]{stinson09} Stinson, G.~S., 
Dalcanton, J.~J., Quinn, T., Gogarten, S.~M., Kaufmann, T., 
\& Wadsley, J.\ 2009, MNRAS, 395, 1455 

\bibitem[T\'oth \& Ostriker(1992)]{toth92} T\'oth, G., Ostriker, J. P. 1992, ApJ, 389, 5

\bibitem[Wechsler et al.(2002)]{wechsler02} Wechsler, R. H., Bullock, J. S., Primack, J. R., Kravtsov, A. V., Dekel, A. 2002, ApJ, 568, 52

\bibitem[Wetterer \& McGraw(1996)]{wetterer96} Wetterer, C., McGraw, J. 1996, AJ, 112, 1046

\bibitem[Wiersma et al.(2009a)]{wiersma09a} Wiersma, R.~P.~C., Schaye, J., Smith, B.~D.  2009a, MNRAS, 393, 99

\bibitem[Wiersma et al.(2009b)]{wiersma09b} Wiersma, R.~P.~C., Schaye, J., Theuns, T., Dalla Vecchia, C., Tornatore, L.  2009b, MNRAS, 399, 574

\bibitem[Vivas \& Zinn(2006)]{vivas06} Vivas, K., Zinn, R. 2006, AJ, 132, 714

\bibitem[Yanny et al.(2000)]{yanny00} Yanny, B. et al. 2000, ApJ, 540, 825

\bibitem[Zibetti et al.(2004)]{zibetti04} Zibetti, S., White, S.~D.~M., Brinkman, J. 2004, MNRAS, 347, 556
 
\bibitem[Zinn(1985)]{zinn85} Zinn, R. 1985, ApJ, 293, 424

\bibitem[Zolotov et al.(2009)]{zolotov09} Zolotov, A., Willman, B., Brooks, A. M., Governato, F., Brook, C. B., Hogg, D. W., Quinn, T., Stinson, G. 2009, ApJ, 702, 1058

\bibitem[Zolotov et al.(2010)]{zolotov10} Zolotov, A., Willman, B., Brooks, A. M., Governato, F., Hogg, D. W., Shen, S., Wadsley, J. 2010, 721, 738

\end{thebibliography}
\end{document}